\documentclass[aps,prb,floatfix,epsfig,twocolumn,showpacs,preprintnumbers]{revtex4}
\UseRawInputEncoding
\usepackage{amsfonts}
\usepackage{amssymb}
\usepackage{graphicx}
\usepackage{amsmath}
\usepackage{tablefootnote}
\usepackage{setspace}

\begin{document}

\title{The relativistic 5f electronic structure of delocalized $\alpha$-U and localized $\delta$-Pu from the self consistent vertex corrected GW approach and X-ray Emission Spectroscopy}

\author{A.~L.~Kutepov$^{1}$\footnote{e-mail: akutepov@bnl.gov}, J.~G.~Tobin$^{2}$, S.-W.~Yu$^{3}$, B.~W.~Chung$^{3}$, P.~Roussel$^{4}$, S.~Nowak$^{5}$, R.~Alonso-Mori$^{5}$, T.~Kroll$^{5}$,  D.~Nordlund$^{5}$,  T.-C.~Weng$^{5}$,  and D.~Sokaras$^{5}$}

\affiliation{$^{1}$Condensed Matter Physics and Materials Science Department, Brookhaven National Laboratory, Upton, NY 11973}
\affiliation{$^{2}$Departments of Physics and Chemistry, University of Wisconsin-Oshkosh, Oshkosh, Wisconsin 54901}
\affiliation{$^{3}$Lawrence Livermore National Laboratory, Livermore, California 94550}
\affiliation{$^{4}$AWE plc, Aldermaston, Reading, Berkshire RG7 4PR, United Kingdom}
\affiliation{$^{5}$SLAC National Accelerator Laboratory, Menlo Park, CA 94025, USA}

\begin{abstract}
The recently developed self-consistent vertex corrected GW method is used to calculate the 5f electronic structure in delocalized $\alpha$-U and localized $\delta$-Pu, each of which is confirmed by the historical experimental approaches of direct and inverse photoemission.  Tender X-Ray Emission Spectroscopy (XES), in a novel application to 5f electronic structure, is used to experimentally prove the existence of 5f delocalization in $\alpha$-U.
\end{abstract}

\maketitle


\section*{Introduction}
\label{intr}

Actinide elements and their compounds play increasingly important role in electric power production. Also, their role in nuclear weapons is well known. Along with actinides' usefulness of being able to produce energy for modern technological society, comes an unwelcome list of their "side effects" such as radioactivity, chemical reactivity, toxicity. Therefore, for their safe and efficient use as energy source, a robust knowledge of actinides' properties is of utmost importance. Unfortunately, our understanding of these fascinating elements and their compounds in many aspects is far from being complete. One such aspect is the structure of occupied and unoccupied electronic states which we will address in this work (in somewhat simplified terms) as ODOS (occupied density of states) and UDOS (unoccupied density of states) correspondingly. Knowledge of the ODOS/UDOS is important for studying the thermodynamic properties as well as for understanding the materials' response when external perturbations such as X-rays or electronic beams are applied. Despite intense experimental\cite{iopmse_9_012054,prb_72_085109,topcat_56_1104,jesrp_194_14,jvsta_39_043205} and theoretical\cite{epl_77_17003,nature_446_513,prb_75_235107,prb_76_245118,prl_101_056403,prl_101_126403,prb_102_245111,prb_101_125123,ncomm_4_2644,prb_101_245156,sadv_1_1,prb_99_125113}, research, a unified picture of ODOS and UDOS in actinides remains elusive. From the experimental point of view, the difficulties come from the above mentioned list of unwelcome properties (radioactivity, chemical reactivity, toxicity) as well as from the lack of adequate theoretical models for interpretation of available experimental data. From the theoretical point of view the difficulties arise from the combined presence of the relativistic and correlation effects in actinide materials.

The principal and fundamental knowledge about materials properties comes from experimental measurements. Concerning the ODOS/UDOS, historically the main experimental technics are Photo-Electron Spectroscopy\cite{prb_65_235118} (PES), Resonance Photo-Electron Spectroscopy\cite{prb_68_155109} (ResPES), Inverse Photo-Electron Spectroscopy\cite{ss_714_121914} (IPES), Bremstralung Isochromate Spectroscopy\cite{pr_6_166,rsi_50_221} (BIS), X-ray absorption spectroscopy\cite{phsr_102_241}(XAS), and the electron energy loss spectroscopy\cite{iopmse_109_012007}(EELS). PES and ResPES directly measure the occupied density of states. IPES and BIS directly measure the unoccupied electronic states. PES/ResPES and IPES/BIS, therefore, can correspondingly be explicitly compared with the calculated ODOS and UDOS. XAS and EELS provide indirect information about UDOS, particularly in those cases where the 5f electronic states are localized. Interpretation of the XAS and EELS data requires a knowledge of UDOS and some additional information such as matrix elements of dipole transitions. Ab-initio theoretical modeling of ODOS and UDOS, therefore, is important not only to assess the quality of different theoretical approaches by comparison them with the corresponding experimental data but also to help with the interpretation of experimental results.

Concerning the experimental methods, one particular limitation of the conventional XAS/EELS approach is an inability to resolve the difference in 5f electronic structure between localized U and delocalized U.\cite{prb_72_085109}  Here, we will present new and powerful X-ray Emission Spectroscopy (XES) measurements that will fill this void.

Theoretical (calculational) knowledge of ODOS/UDOS in actinides is based mostly on the results of calculations, [\onlinecite{prb_55_1997,prb_67_235105,prb_72_085109}], performed within the density functional theory assuming local density approximation (LDA) or the generalized gradient approximation (GGA) for the exchange and correlation effects. Whereas such calculations are computationally inexpensive, the underlying theory is designed for the ground state properties, not for the excited states. Plus, both LDA and GGA are supersimplified approximations for the exchange-correlation effects and, therefore, can hardly be considered as adequate when applied to the f-electron systems. Another issue which precludes application of many existing band structure codes to actinides is the importance of relativistic effects. With an exclusion of just a few codes, these effects are normally treated in the scalar-relativistic (SR) approximation\cite{zpb_32_43} or, at best, by treating spin-orbit interaction perturbatively/variationally with a basis set constructed from SR functions. Such approximations are insufficient for actinides.\cite{prb_63_035103,as_9_5020}

Another calculational approach, which has been intensely used to study actinides lately, is the combination of LDA and dynamical mean field theory (LDA+DMFT, [\onlinecite{rmp_68_13}]). In essence, LDA+DMFT replaces LDA exchange-correlation effects in f-shell of electrons locally (onsite) with the ones resulted from the so called local impurity model with parametrically given interactions. Despite being semi-empirical, LDA+DMFT is very popular in applications to actinides. Unfortunately, the parametric nature of the LDA+DMFT approach makes it difficult to provide unique ODOS/UDOS. In particular, electronic structure of $\delta$-Pu which was intensily studied during the last two decades shows considerable differences from one publication to another. For instance, in Refs. [\onlinecite{prb_102_245111,prb_101_245156,prl_101_126403}], DOS of $\delta$-Pu shows a single-peak structure (at $E_{F}$) and rather flat featuresless UDOS. At the same time, in Refs. [\onlinecite{epl_77_17003,prb_101_125123,prb_75_235107}], UDOS of $\delta$-Pu demonstrates the peaks but at energies 4 eV or higher which severely overestimates the experimental observations (see below). There are also LDA+DMFT publications\cite{nature_446_513,prb_76_245118} where the peak of UDOS is situated at about 2 eV above the Fermi level which is in accordance with our theoretical and experimental observations (see below). The reasons for such diversity of the LDA+DMFT results, actually, are many. Let us summarize them here for the purpose of stimulating the discussions. Probably the most serious reason is the fact that "beyond LDA" exchange-correlation DMFT effects are embedded in extremely small subspace: onsite-only and f-electrons-only. The failure of onsite-like approximations was recently investigated for the "beyond GW" diagrams in a few materials\cite{prm_5_083805,jcm_33_485601}, including one f-electron material. In Refs. [\onlinecite{prm_5_083805,jcm_33_485601}] the analysis was in the context of GW+DMFT approach. For LDA+DMFT, the issue of too small correlated subspace should be even more serious, because the approximation of locality in LDA+DMFT affects not only "beyond GW" diagrams but also the GW diagram itself (as a part of DMFT correction). Another issue, related to the local approximations, is the ambiguity of how to actually define the local subspace (choice of local orbitals and projectors). Karp et al., [\onlinecite{prb_103_195101}], have recently shown that LDA+DMFT results can be seriously affected by different choices of the local subspace. The next issue is the ambiguity with the double-counting:  interaction between electrons in the onsite f-shell is counted twice - in LDA and in DMFT. The correction of this fact is not well defined in theory and it is not unique. Finally, interactions in the correlated subspace are not defined self-consistently but instead are given manually by the Hubbard parameters U and J. Parameters U and J in LDA+DMFT calculations normally are adjusted to ensure correct ground state properties but not the electronic structure (particularly not UDOS) and, therefore, the obtained UDOS in LDA+DMFT calculations differs a lot from one publication to another reflecting the choice of U and J and also the mentioned ambiguities in the definition of local subspace and in double-counting corrections. The choice of the so called solver of the impurity problem also contributes to the differences.

Applications of ab-initio approaches which go beyond DFT systematically (i.e. which are free from "local" approximations and of adjustable parameters) for modeling the actinides are very scarce yet. In the work by Chantis et al., [\onlinecite{prb_78_081101}], quasi-particle self consistent GW method (QSGW) was applied to study electronic structure of uranium. Calculations were performed in scalar-relativistic approximation with spin-orbit interaction evaluated perturbatively. It was shown that QSGW approximation predicts an f-level shift upward of about 0.5 eV with respect to the s−d states and that there is a f-band narrowing when compared to LDA band-structure results. A few years ago we used scGW (fully self consistent GW) approach to study electronic structure of Pu and Am metals\cite{prb_85_155129} and found remarkable improvement in description of occupied DOS of americium as compared to the results obtained with DFT or QSGW. However, the emphasis in those scGW calculations was mostly on the occupied DOS, not on the UDOS. Also, $\alpha$-U was not studied in Ref. [\onlinecite{prb_85_155129}]. Applications of ab-initio approaches which go diagrammatically beyond GW (i.e. include vertex corrections) in the physics of actinides are still not existent.

This work, therefore, has several goals. The first goal is to compare very recent IPES results for $\delta$-Pu UDOS with newly developed self-consistent vertex-corrected GW method (sc(GW+G3W2), Ref. [\onlinecite{prb_94_155101}]) as well as with scGW and LDA results. The second goal is to compare very recent IPES for $\alpha$-U with older BIS data and with theoretical calculations for this material. Here we believe that theoretical calculations can help to judge the pros and cons of IPES/BIS which have already been discussed by Paul Roussel et al. in Ref. [\onlinecite{ss_714_121914}] but without using a theoretical information. The third goal is to compare theoretical calculations of ODOS with experimental PES/ResPES results. The fourth goal is to compare three theoretical approximations (LDA, scGW, and sc(GW+G3W2)) between each other in order to assess the quality of them in application to selected actinide elements and to estimate the strength of correlation effects in them. A fifth goal is to demonstrate that U XES can easily differentiate between 5f localized behavior and the 5f delocalized behavior of $\alpha$-U, providing a new and powerful way to quantify 5f delocalization.

The paper begins with two sections presenting theoretical and experimental approaches which we use in this study. Then we present the results of our theoretical calculations and compare them with experimental PES/ResPES and BIS/IPES data including discussion. Next goes a section devoted to a new direction of the research in the field of ODOS/UDOS of actinides, X-ray Emission Spectroscopy, where first new results for $\alpha$-U and uranium compounds are presented. Finally, we draw conclusions.

\section*{Methods and calculation setups}\label{meth}

All calculations in this work were performed using code FlapwMBPT.\cite{flapwmbpt_2} Recently, a few updates were implemented in the code.\cite{prb_103_165101,jcm_33_235503} For DFT calculations, we used the local density approximation (LDA) as parametrized by Perdew and Wang.\cite{prb_45_13244} We also used two diagrammatic approaches: scGW (fully self-consistent GW) and sc(GW+G3W2) (fully self-consistent GW plus vertex correction of first order). These two diagrammatic methods are based on the Hedin equations.\cite{pr_139_A796} For convenience, we remind the reader about how Hedin's equations could be solved self-consistently in practice.

Suppose one has a certain initial approach for Green's function $G$ and screened interaction $W$. Then one calculates the following quantities:

three-point vertex function from the Bethe-Salpeter equation

\begin{align}\label{Vert_0}
\Gamma^{\alpha}(123)&=\delta(12)\delta(13)\nonumber\\&+\sum_{\beta}\frac{\delta \Sigma^{\alpha}(12)}{\delta
G^{\beta}(45)}G^{\beta}(46)\Gamma^{\beta}(673) G^{\beta}(75),
\end{align}
where $\alpha$ and $\beta$ are spin indexes, and the digits in the brackets represent space-Matsubara's time arguments,

polarizability

\begin{equation}\label{def_pol1}
P(12)=\sum_{\alpha}G^{\alpha}(13)\Gamma^{\alpha}(342)G^{\alpha}(41),
\end{equation}

screened interaction

\begin{align}\label{def_W2}
W(12)=V(12) +V(13)P(34)W(42),
\end{align}

and the self energy

\begin{equation}\label{def_M8}
\Sigma^{\alpha}(12)= - G^{\alpha}(14)\Gamma^{\alpha}(425)W(51).
\end{equation}

In the equation (\ref{def_W2}) V stands for the bare Coulomb interaction.
New approximation for the Green function is obtained from Dyson's equation

\begin{align}\label{D4}
G^{\alpha}(12)=G_{0}^{\alpha}(12) +G_{0}^{\alpha}(13)\Sigma^{\alpha}(34)G^{\alpha}(42),
\end{align}
where $G_{0}$ is the Green function in Hartree approximation. Eqn. (\ref{Vert_0}-\ref{D4}) comprise one iteration. If convergence is not yet reached one can go back to the equation (\ref{Vert_0}) to
start the next iteration with renewed $G$ and $W$.

\begin{figure}[t]    
    \fbox{\includegraphics[width=3.6 cm]{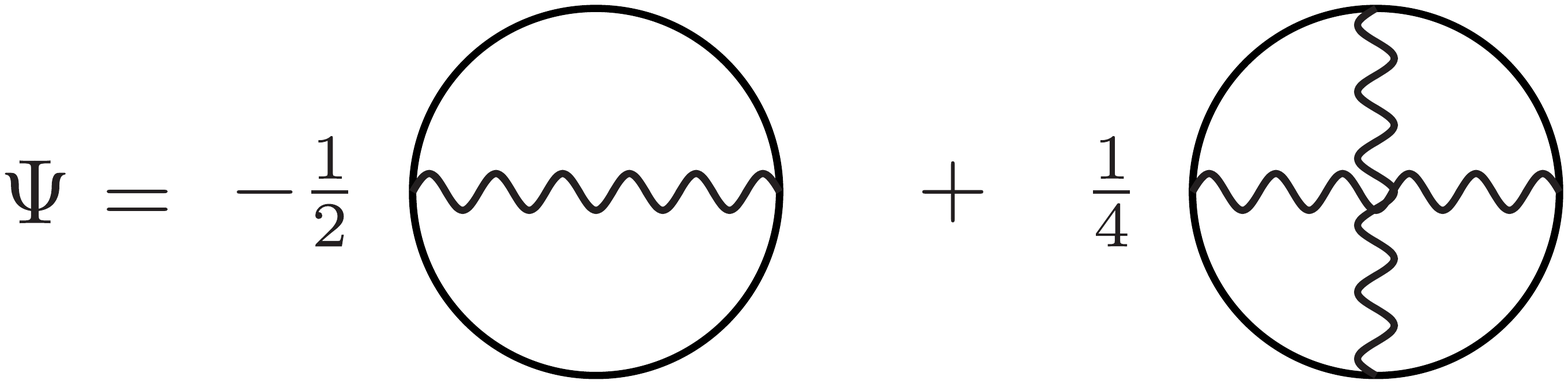}}   
    \hspace{0.02 cm}
    \fbox{\includegraphics[width=3.6 cm]{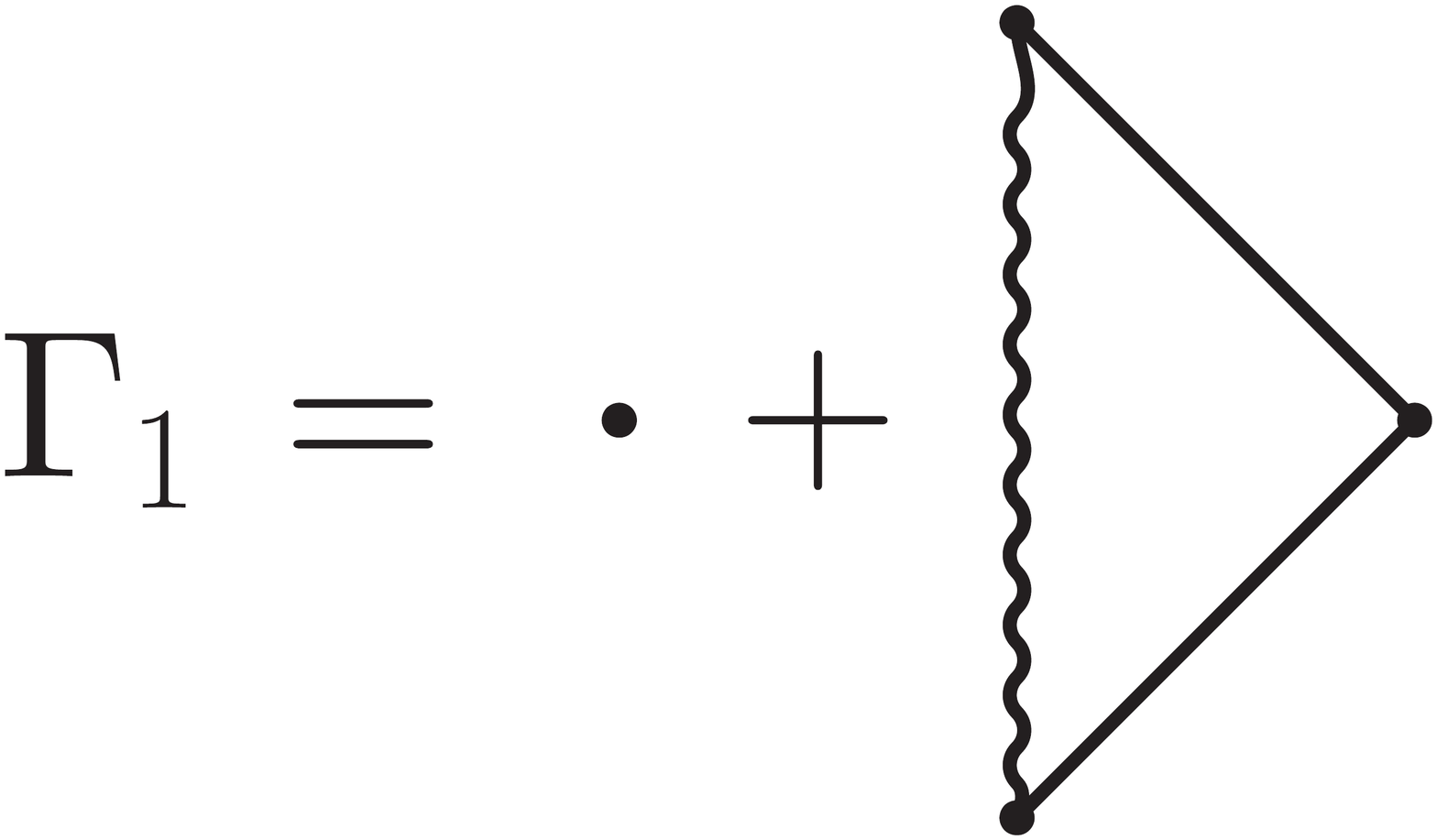}}  
    \hspace{0.02 cm}
    \fbox{\includegraphics[width=3.6 cm]{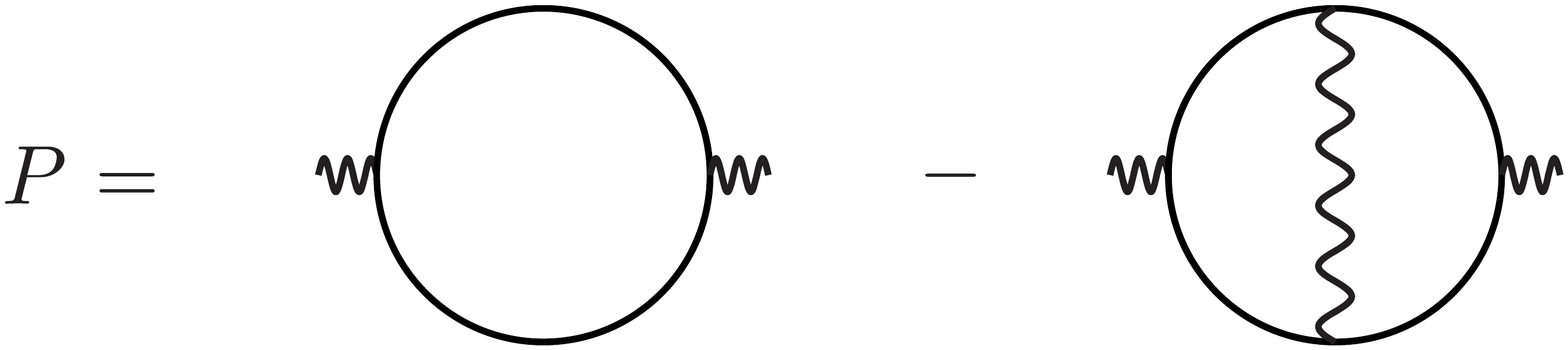}}  
    \hspace{0.02 cm}
    \fbox{\includegraphics[width=3.6 cm]{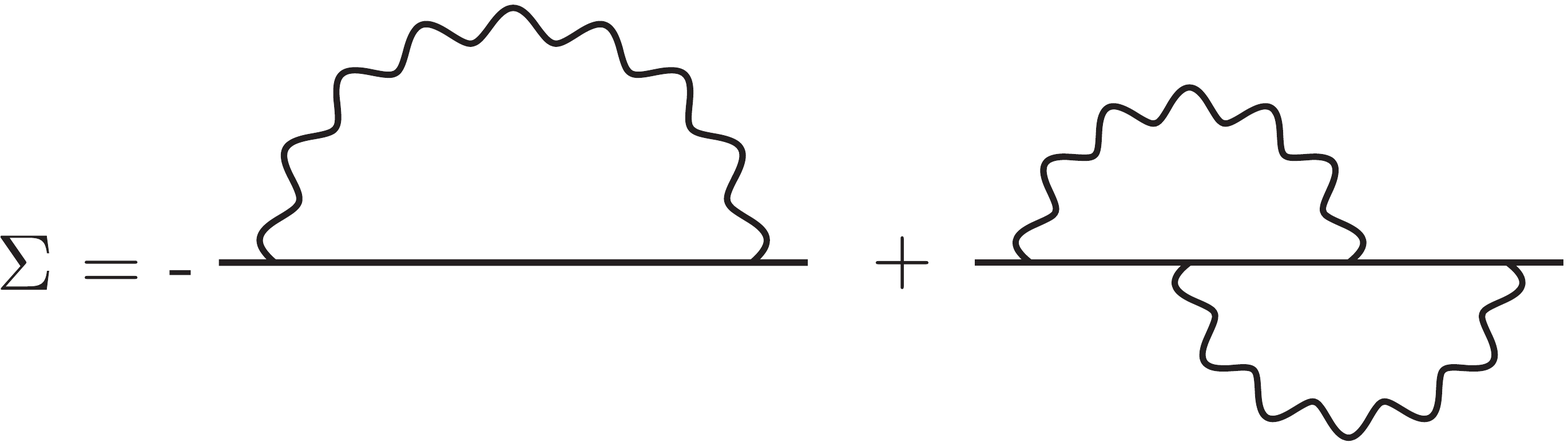}}
\caption{Diagrammatic representation of sc(GW+G3W2) approximation. Top left: $\Psi$-functional. Top right: three-point vertex function. Bottom left: polarizability. Bottom right: self energy. scGW approximation corresponds to neglecting the second term in all figures.}
\label{sigma}
\end{figure}

The system of Hedin's equations formally is exact, but one has to introduce certain approximations for the vertex function $\Gamma^{\alpha}(123)$ in order to make the solving of the system manageable in practice. Approximations for the vertex function which we use in this study are presented diagrammatically in Fig. \ref{sigma} (top right). First term (trivial vertex) corresponds to scGW and the addition of the first order vertex corresponds to sc(GW+G3W2) approximation. Substitution of the selected vertex function in equations (\ref{def_pol1}) and (\ref{def_M8}) specifies the corresponding approximations for polarizability and self energy which also are shown in Fig. \ref{sigma}. Diagrammatic approaches which are used in this work, scGW and sc(GW+G3W2), can also be defined using $\Psi$-functional formalism of Almbladh et al.\cite{ijmpb_13_535} Corresponding $\Psi$-functional which includes vertex corrections up to the first order in screened interaction W is shown in Fig. \ref{sigma} (top left). As in the case of vertex function, the first diagram corresponds to GW approximation, whereas the sum of the first and the second diagram represents sc(GW+G3W2) approximation. Diagrammatic representations for irreducible polarizability (Fig. \ref{sigma}, bottom left) and for self energy (Fig. \ref{sigma}, bottom right)  follow from the chosen approximation for $\Psi$-functional: $P=-2(\frac{\delta \Psi}{\delta W})_{G}$ and $\Sigma=(\frac{\delta \Psi}{\delta G})_{W}$.

\begin{figure}[t]
\begin{center}       
\includegraphics[width=8.5 cm]{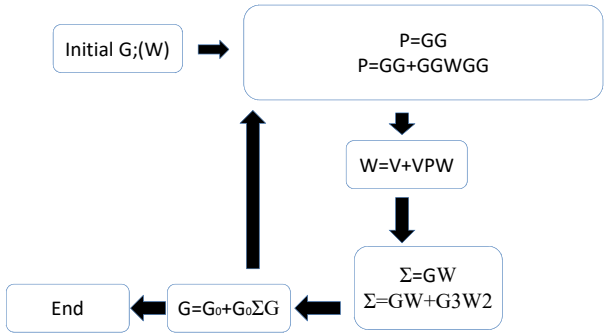}
\caption{Flowchart of scGW and sc(GW+G3W2) calculations. All equations are presented using symbolic notations. In the expressions for polarizability and self energy, first equation corresponds to scGW, second equation is used in sc(GW+G3W2). $G_{0}$ stands for Green's function in Hartree approximation. Any calculation begins with self-consistent DFT iterations where the basis set is formed and the initial approach for G is generated. Iterations of scGW method use this initial Green's function as an input in order to start. During scGW iterations, G is updated and screened interaction W is generated. Both G and W serve as an input to start iterations of sc(GW+G3W2) approach. Not shown is that $G_{0}$ depends on the electronic density (i.e. on $G$) from the previous iteration and is updated in the course of iterations.}
\label{flow}
\end{center}
\end{figure}

In practice, every calculation begins with self-consistent DFT step which generates the band states which are used as a basis set for subsequent diagrammatic approaches, scGW and sc(GW+G3W2). The flowchart of scGW/sc(GW+G3W2) is shown schematically in Fig. \ref{flow}. Technical details of the DFT part have been described in Refs. [\onlinecite{jcm_15_2607,prb_103_165101}] and the details of scGW implementation in Refs. [\onlinecite{prb_85_155129,cpc_219_407}]. Detailed account of the algorithms for sc(GW+G3W2) and also for other vertex corrected schemes can be found in Refs. [\onlinecite{prb_94_155101,prb_95_195120,prb_96_035108,prb_104_085109}]. All three methods applied in this study are based on Dirac equation and, therefore, include all important relativistic effects systematically instead of common perturbative/variational treatment of spin-orbit interaction in many other codes.

In order to make presentation more compact, principal structural parameters for the studied solids have been collected in Table \ref{list_s} and the most important set up parameters have been collected in Table \ref{setup_s}. All calculations have been performed for the electronic temperature $600K$. The calculations (excluding the vertex part) were performed with the $12\times 12\times 12$ mesh of \textbf{k}-points in the Brillouin zone ($\delta$-Pu) and $9\times 9\times 6$ mesh ($\alpha$-U). Approximately 220 band states (per atom in the unit cell) were used to expand Green's function and self energy. The diagrams beyond GW approximation were evaluated using $3\times 3\times 3$ mesh of \textbf{k}-points in the Brillouin zone ($\delta$-Pu) and $3\times 3\times 2$ mesh ($\alpha$-U). Approximately 40 bands/atom (closest to the Fermi level) have been used to evaluate higher order diagrams.

\begin{table}[b]
\caption{Structural parameters of the solids considered in this work. Lattice parameters are in Angstroms, muffin-tin (MT) radii are in atomic units (1 Bohr radius), and the atomic positions are relative to the three primitive translation vectors.} \label{list_s}
\small
\begin{center}
\begin{tabular}{@{}c c c c c c c} &Space&&&&Atomic&\\
Solid &group&a&b&c&positions&$R_{MT}$\\
\hline\hline
$\alpha$-U&63 &2.854 &5.869 &4.955&0;0.1025;0.25  &2.6023\\
$\delta$-Pu&225 &4.6347 & && 0;0;0&3.0965\\
\end{tabular}
\end{center}
\end{table}

\section*{Experimental}\label{expr}

The X-Ray Emission Spectroscopy experiments were performed on Beamline 6-2a at the Stanford Synchrotron Radiation Lightsource (SSRL).  These were carried out utilizing both (1) input photons from a Si (111) monochromator and (2) a photon detector, a high-resolution Johansson-type spectrometer, \cite{rsi_91_033101,jesrp_232_100} operating in the tender x-ray regime (1.5 - 4.5 keV).  For the UO$_{2}$ M$_{5}$, U metal M$_{5}$, UF$_{4}$ M$_{5}$ and UF$_{4}$ M$_{4}$ experiments, the excitation photon energies were respectively 3640 eV, 3640 eV, 3650 eV and 3820 eV; each chosen to be significantly above threshold for the transition under consideration.  Instrumentally, the total energy bandpass of this experiment is about 1 eV.  However, the lifetime broadening of the 3d core holes (several eV) dominates the spectral widths. Two layers of Kapton encapsulation were used and data collection times were in the range of a couple of hours. The samples used were the same as used in earlier studies.\cite{jesrp_232_100,ss_698_121607,jpc_4_015013} Uranium samples can be affected by oxidation and sample corruption, but these were not a problem for the UF$_{4}$ and UO$_{2}$ samples, as described earlier,\cite{ss_698_121607,jpc_4_015013} and the correction for the U metal sample is discussed below.  The details of the photoemission (PES, RESPES) and inverse photoemission (IPES, BIS) can be found in the references listed in the text.

\section*{Computation Results and Comparison to Photoemission and Inverse Photoemission}
\label{res}

\begin{table}[t]
\caption{Principal setup parameters of the studied solids are given. The following abbreviations are introduced: $\Psi$ is for wave functions, $\rho$ is for the electronic density, $V$ is for Kohn-Sham potential, and PB is for the product basis.} \label{setup_s}
\small
\begin{center}
\begin{tabular}{@{}c c c c c c} &Core&&$L_{max}$&$L_{max}$&\\
Solid &states&Semicore&$\Psi/\rho,V$&PB & $RK_{max}$ \\
\hline\hline
$\alpha$-U&[Xe]5s,5p,4f&6s,6p,5d&6/6&6&9.0  \\
$\delta$-Pu&[Xe]5s,5p,5d,4f&6s,6p&6/6&6&9.0  \\
\end{tabular}
\end{center}
\end{table}

\begin{figure}[t]    
    \fbox{\includegraphics[width=3.6 cm]{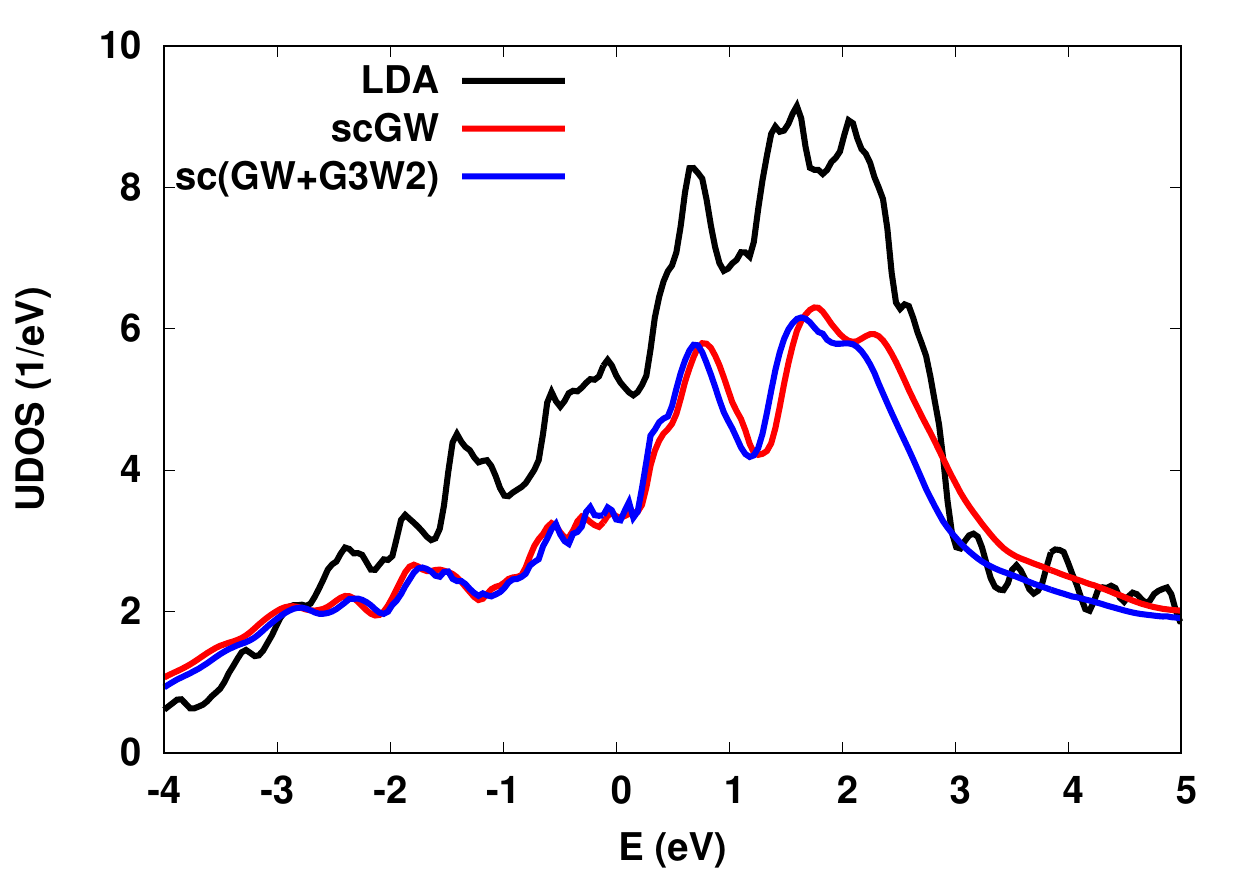}}  
    \hspace{0.02 cm}
    \fbox{\includegraphics[width=3.6 cm]{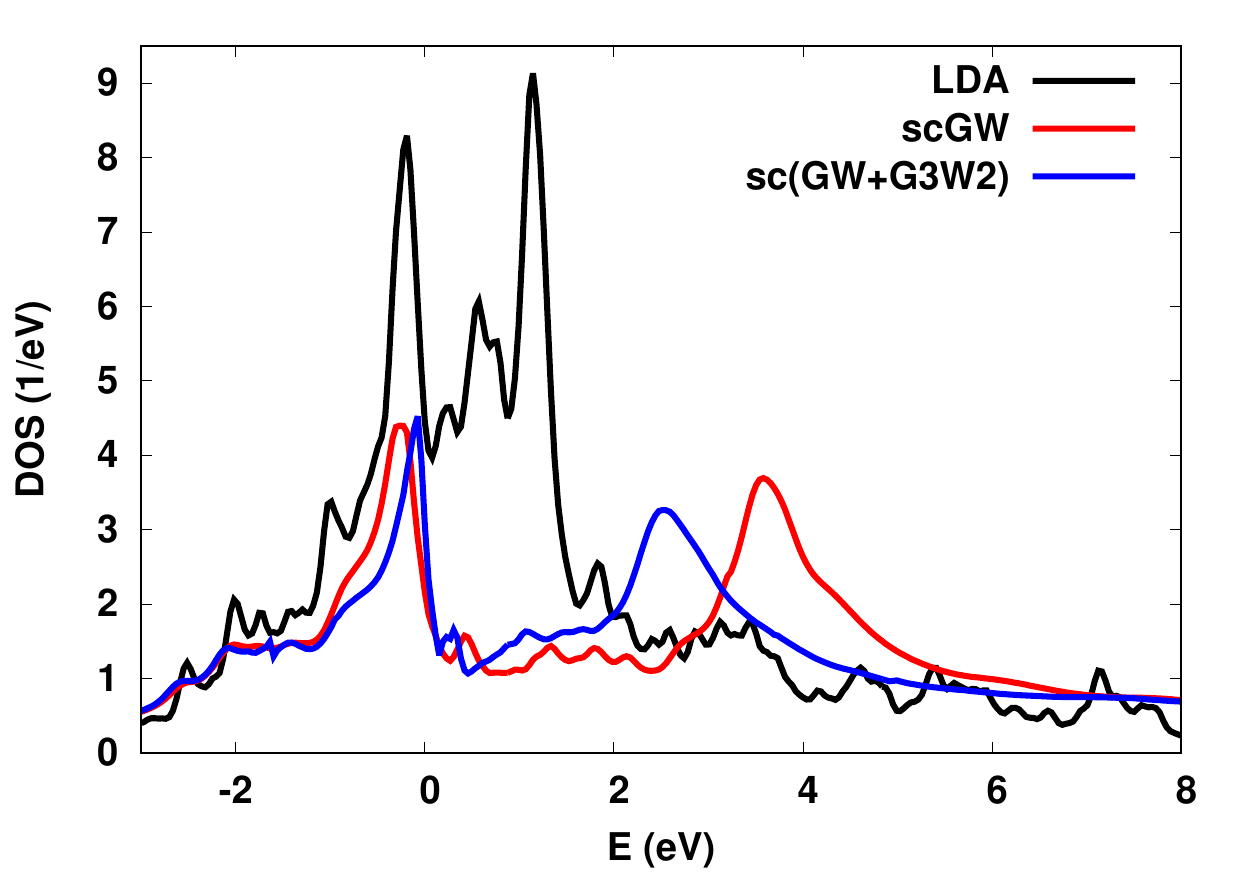}}
\caption{Calculated DOS of $\alpha$-U (left) and $\delta$-Pu (right).}
\label{DOS_theor}
\end{figure}

\begin{figure}[b]    
    \fbox{\includegraphics[width=3.6 cm]{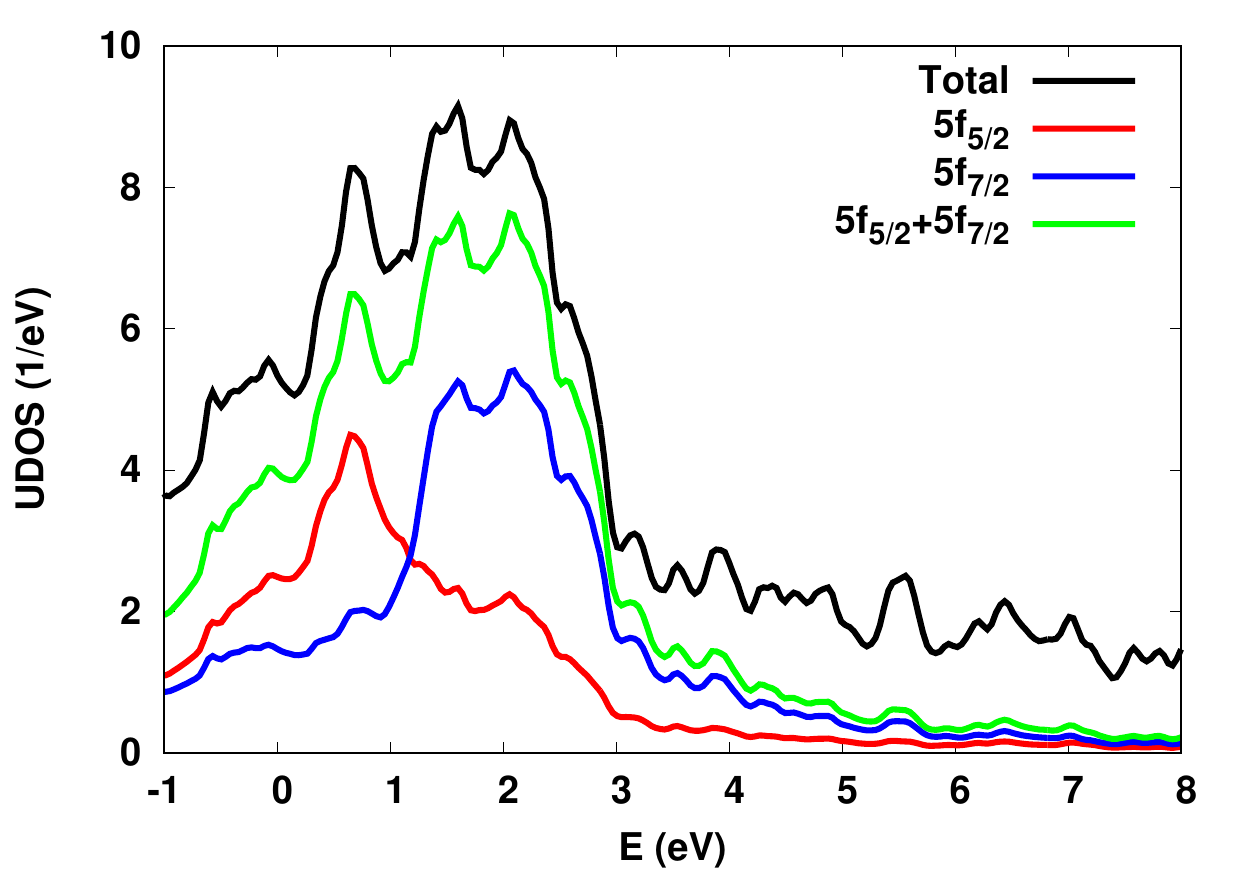}}  
    \hspace{0.02 cm}    
    \fbox{\includegraphics[width=3.6 cm]{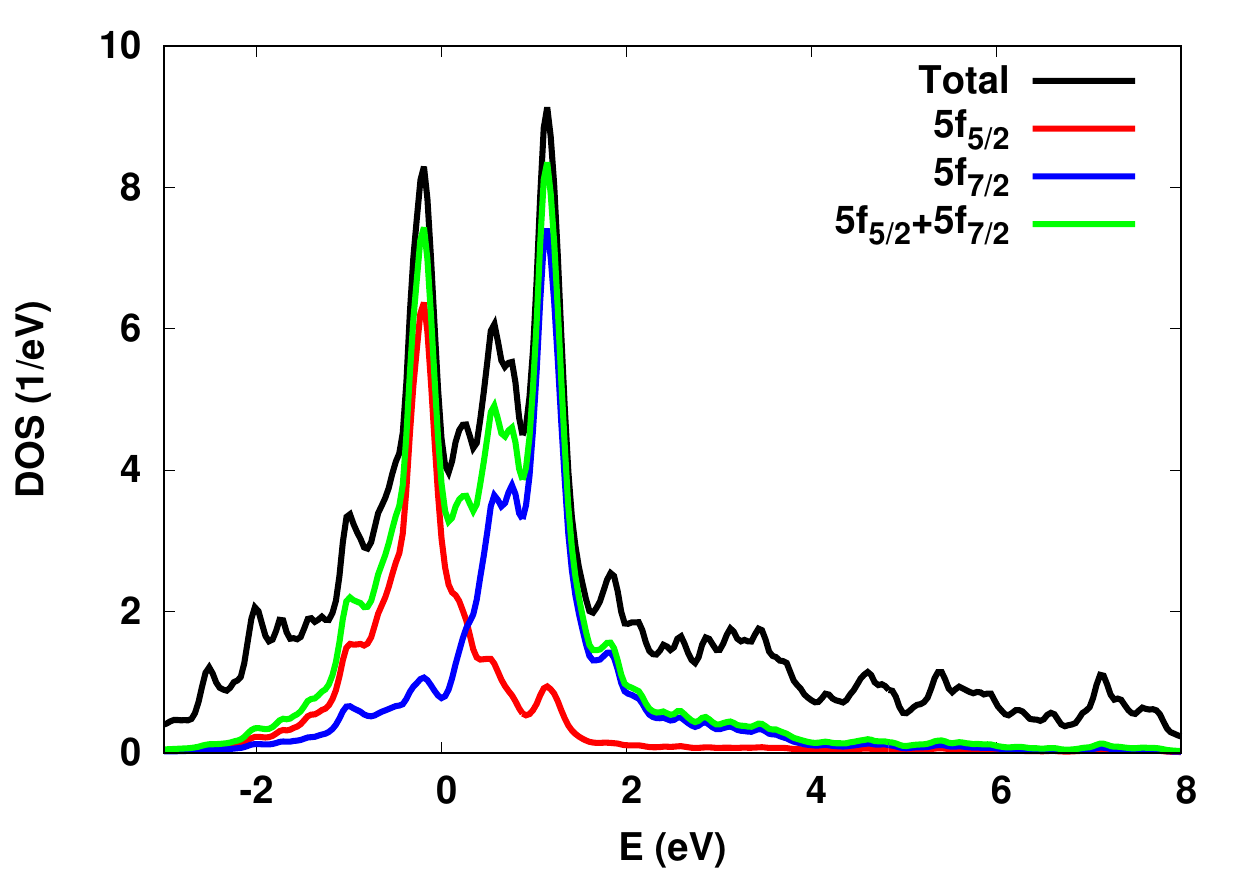}}  
    \hspace{0.02 cm}    
    \fbox{\includegraphics[width=3.6 cm]{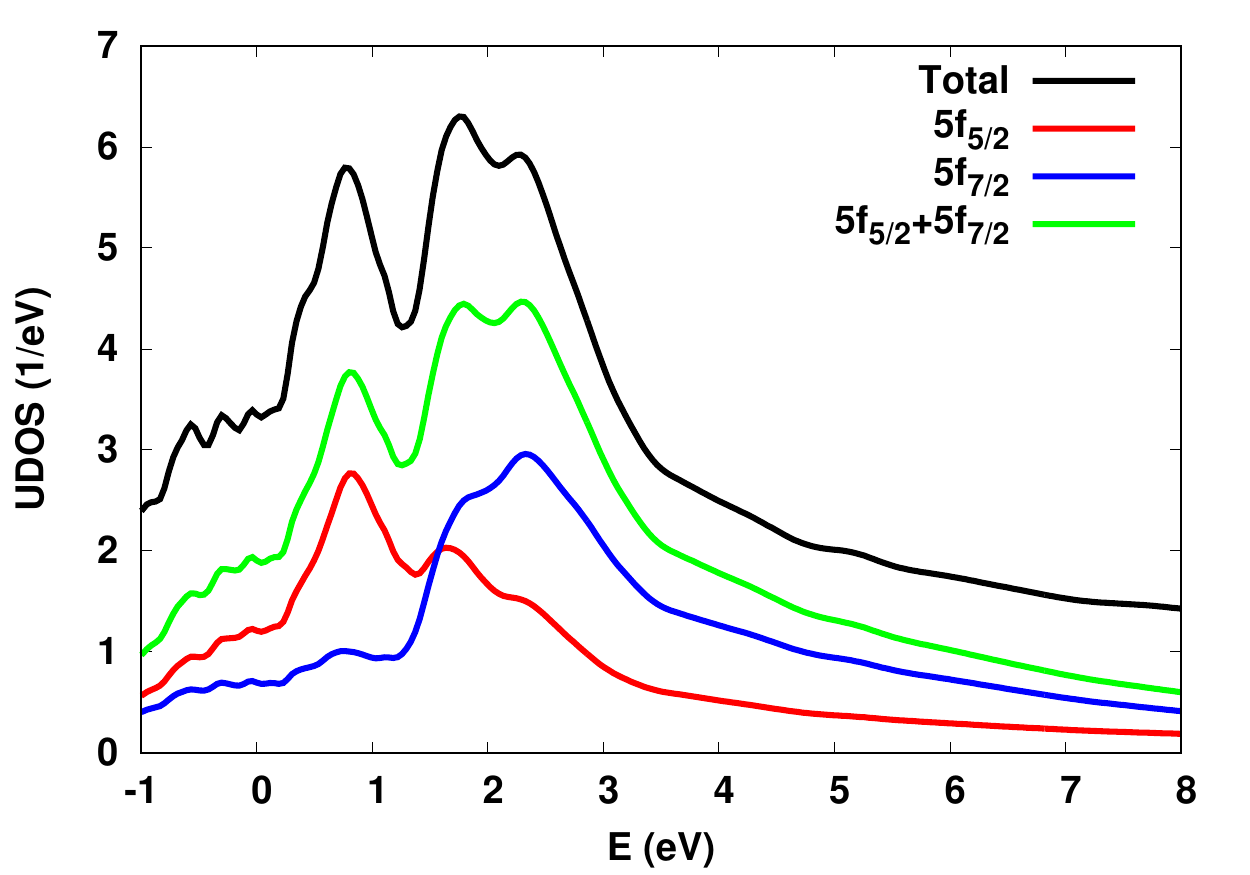}}  
    \hspace{0.02 cm}    
    \fbox{\includegraphics[width=3.6 cm]{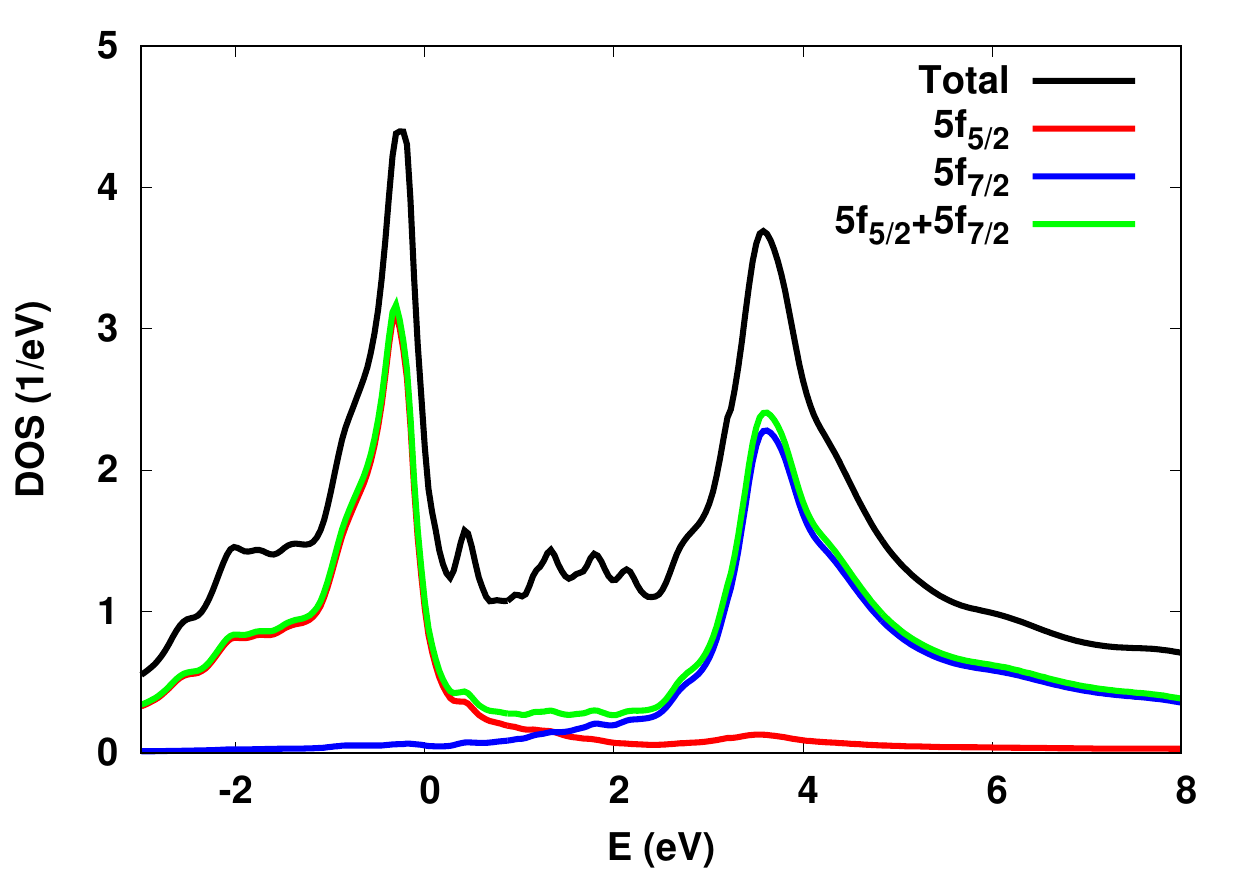}}  
    \hspace{0.02 cm}    
    \fbox{\includegraphics[width=3.6 cm]{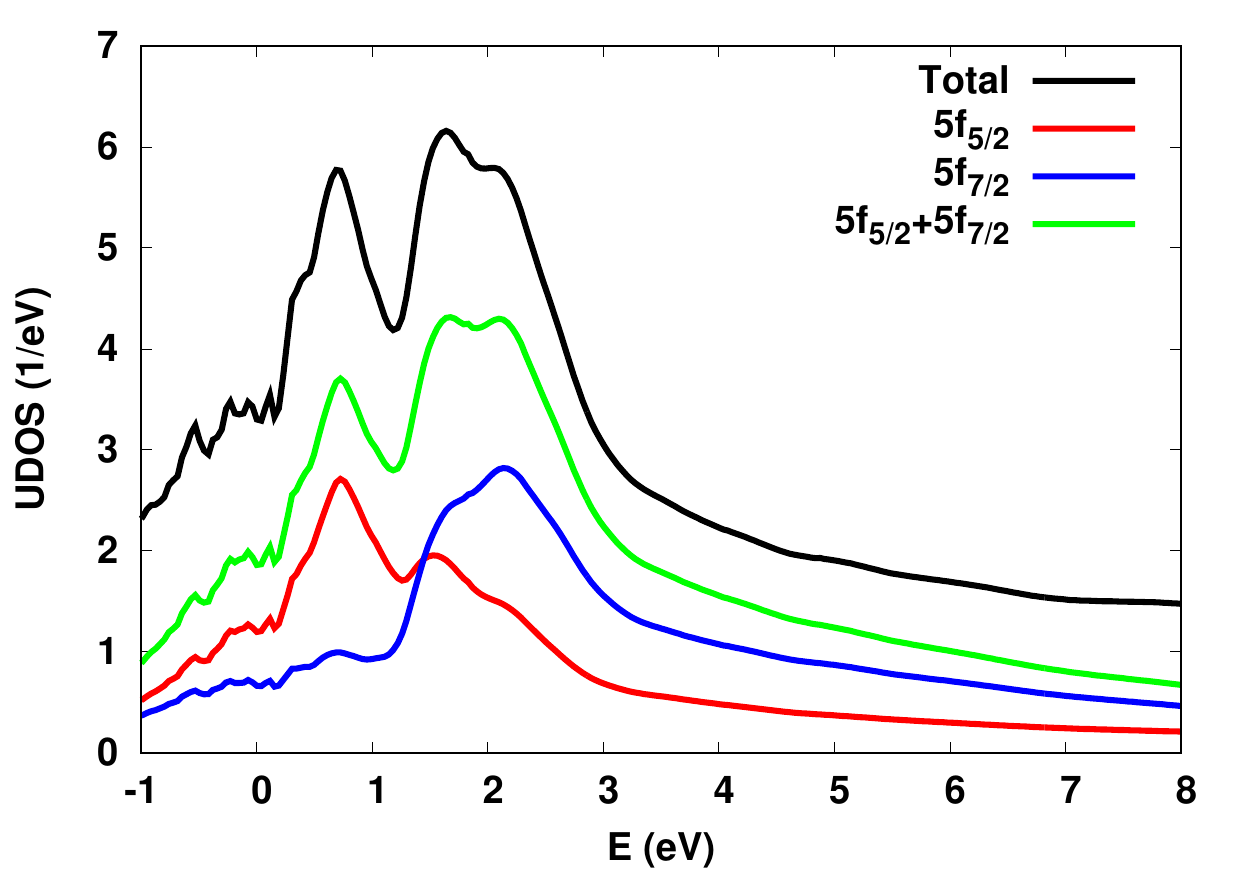}}  
    \hspace{0.02 cm}
    \fbox{\includegraphics[width=3.6 cm]{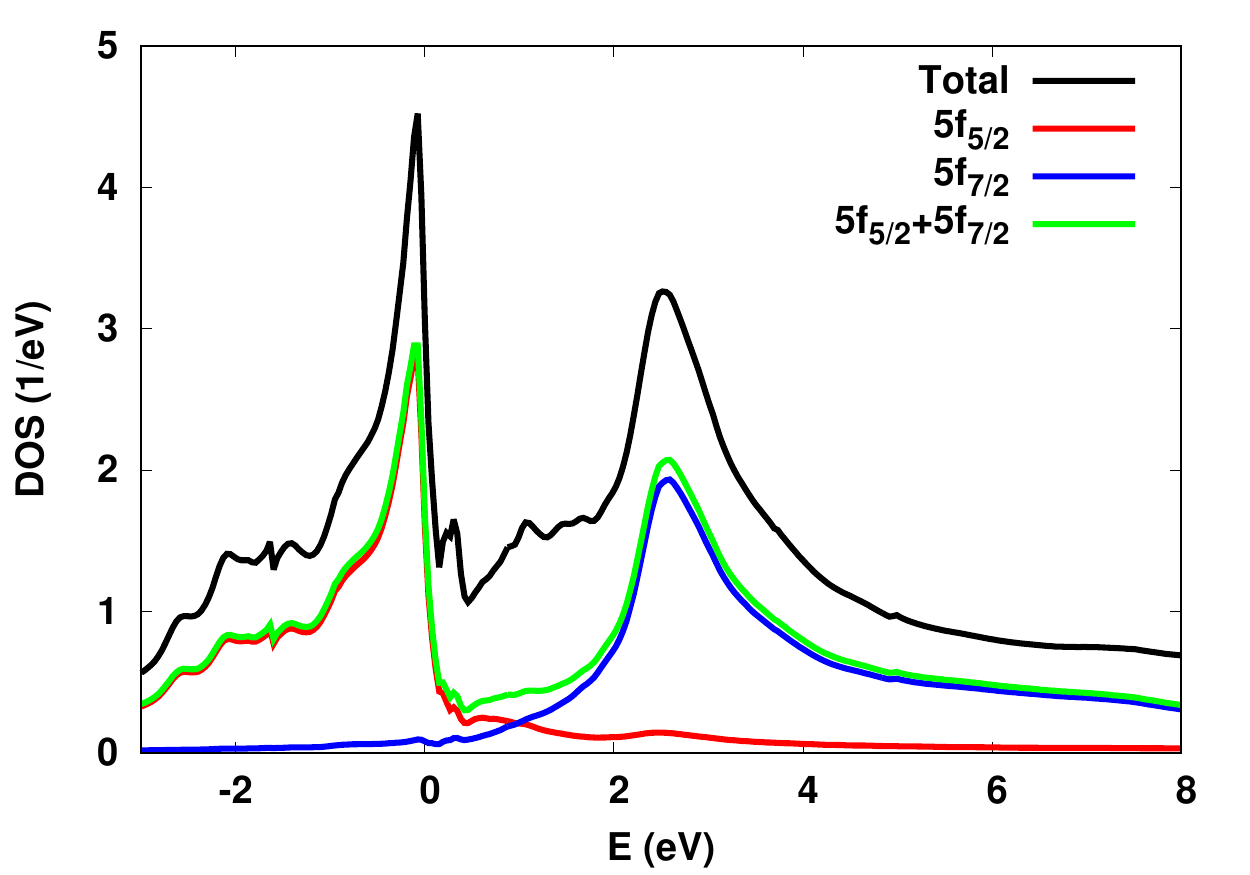}}
\caption{Calculated PDOS of $\alpha$-U (left column) and $\delta$-Pu (right column). In each column, first row corresponds to LDA calculation, second row corresponds to scGW result, and the third row corresponds to sc(GW+G3W2) PDOS.}
\label{PDOS_theor}
\end{figure}

\begin{figure}[t]    
    \fbox{\includegraphics[width=3.6 cm]{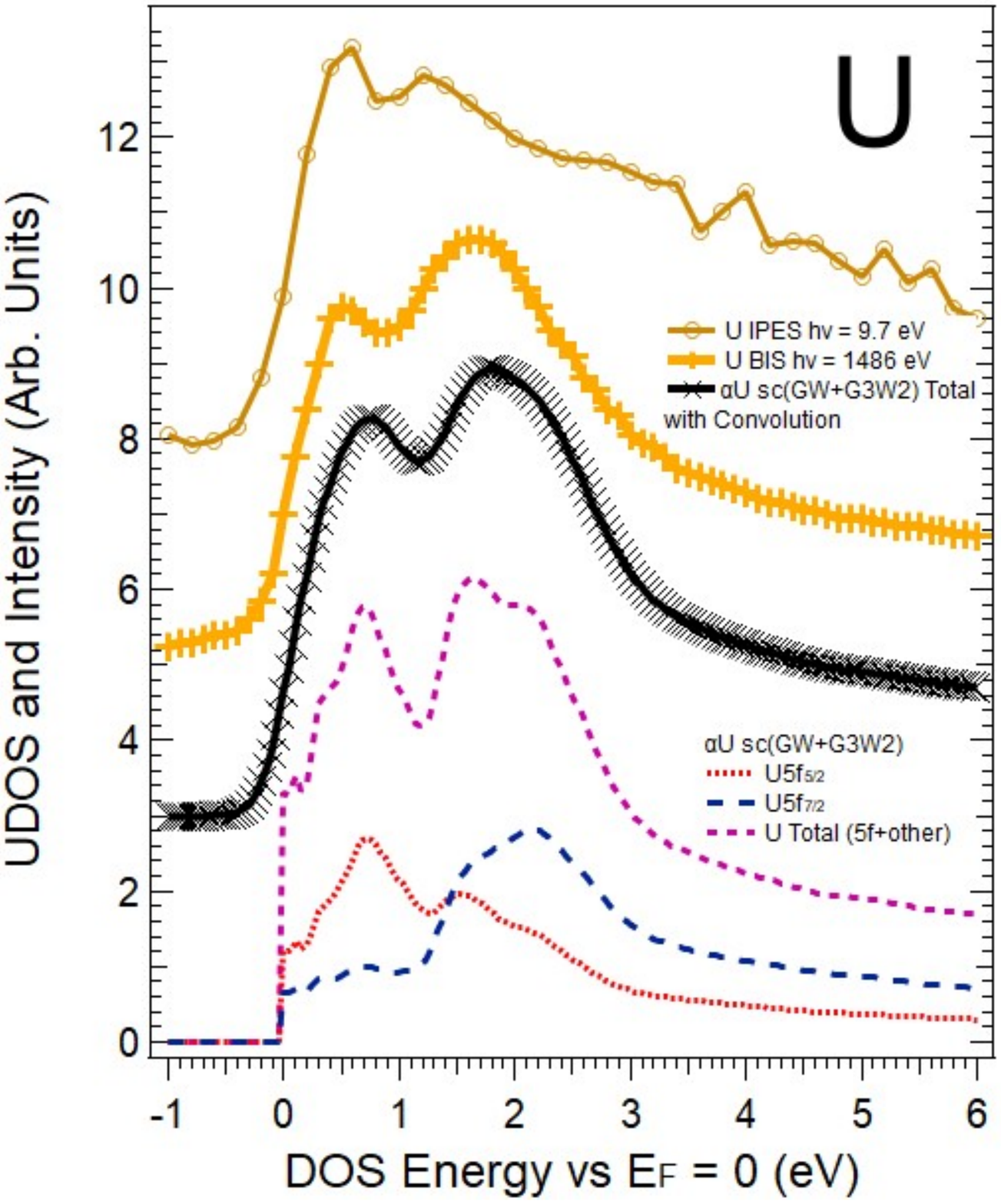}} 
    \hspace{0.02 cm}
    \fbox{\includegraphics[width=3.6 cm]{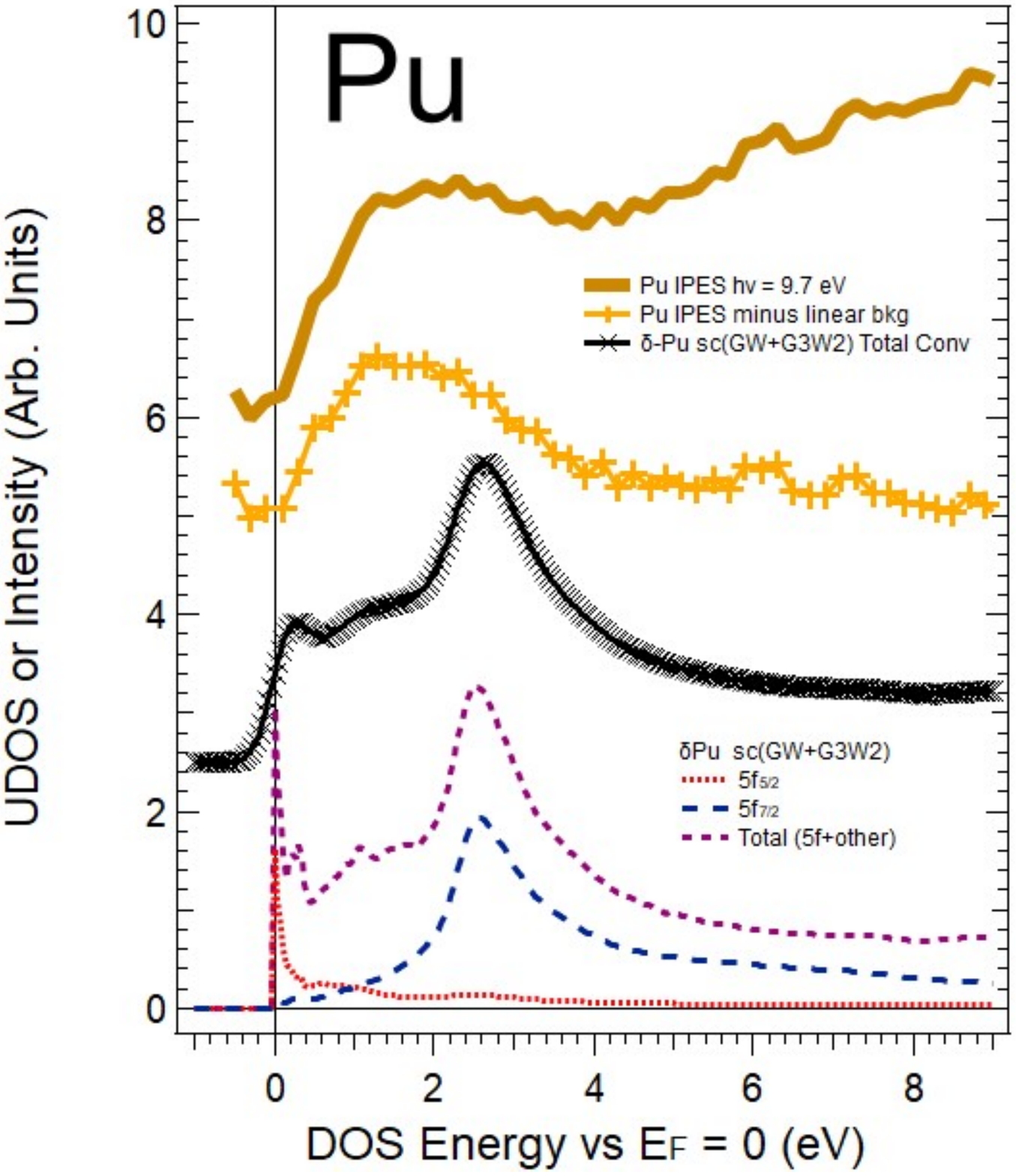}} 
\caption{UDOS of $\alpha$-U (left) and $\delta$-Pu (right). Comparison of BIS ($\alpha$-U only), IPES and theoretical results obtained in sc(GW+G3W2) approximation. The U IPES spectrum has had a linear background subtracted. The Pu IPES spectra have been shifted 0.25 eV. The Total Theory Curves (black x) have been convoluted with a gaussian of 1/2 eV Full Width at Half maximum (FWHM). Some curves have been shifted vertically to enhance the comparison.}
\label{UDOS}
\end{figure}

\begin{figure}[b]    
    \fbox{\includegraphics[width=3.6 cm]{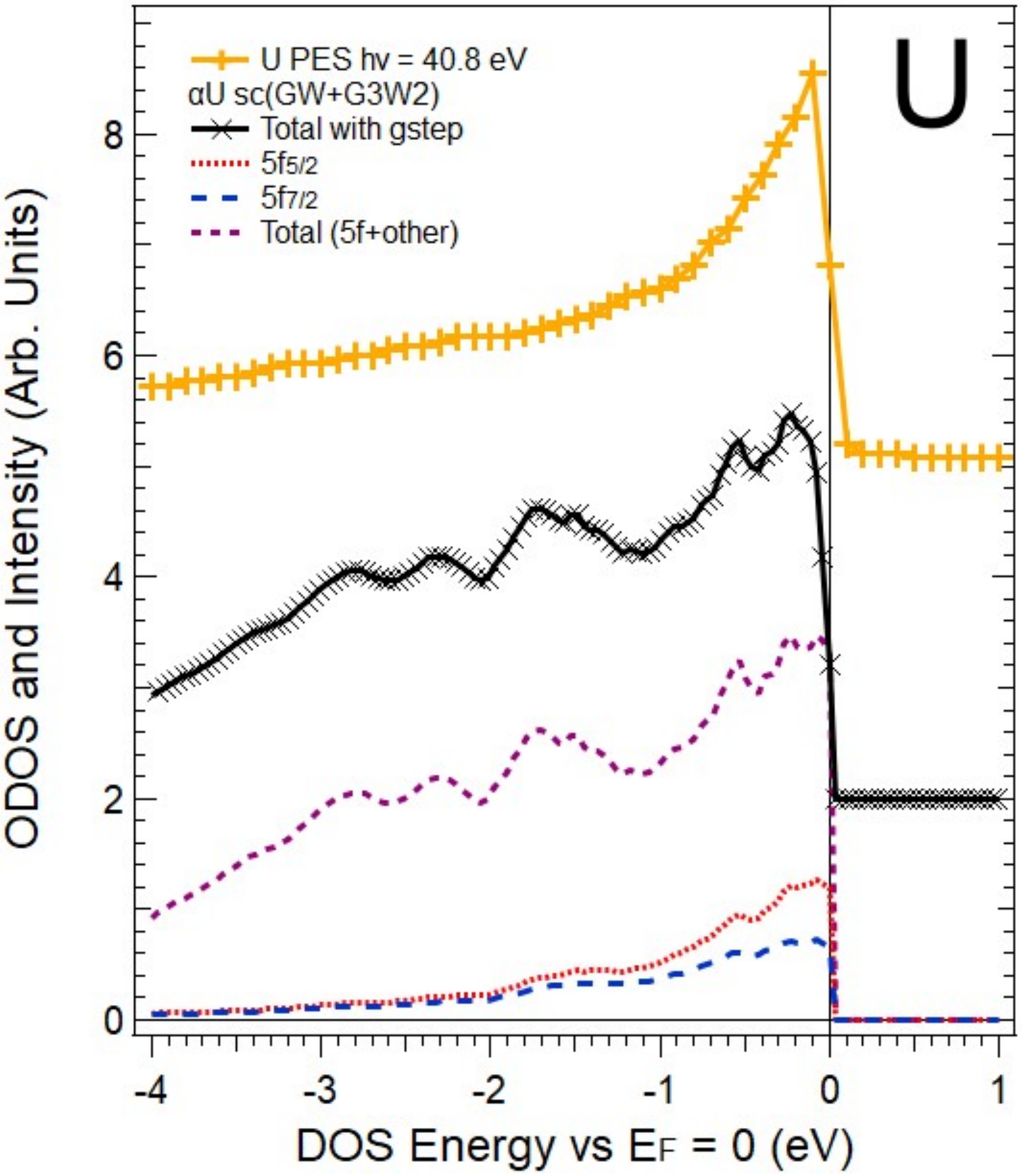}} 
    \hspace{0.02 cm}
    \fbox{\includegraphics[width=3.6 cm]{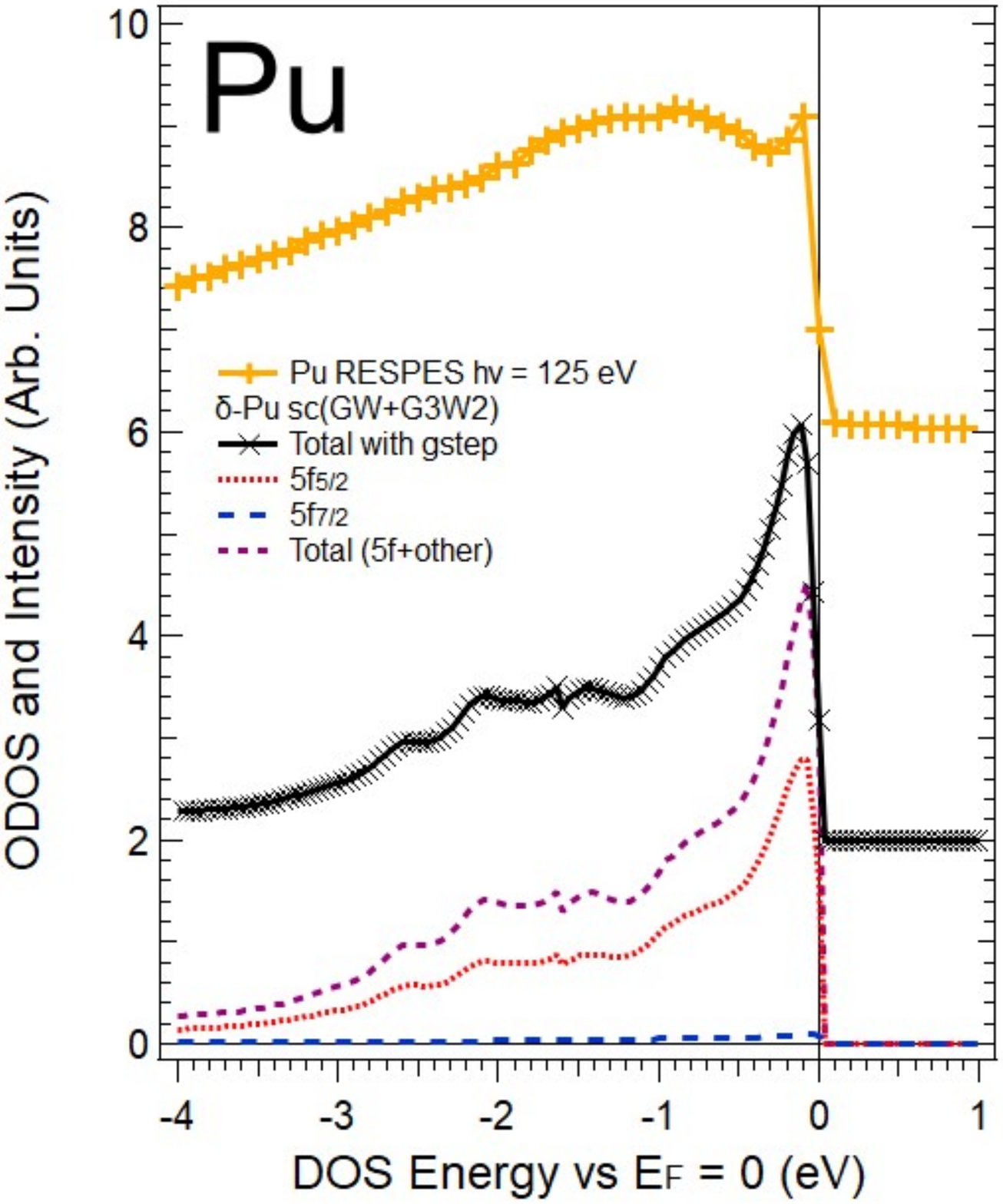}} 
\caption{ODOS of $\alpha$-U (left) and $\delta$-Pu (right). Comparison of ResPES and theoretical results obtained in sc(GW+G3W2) approximation. The Total Theory Curves (black x) have been multipled by a gaussian step function, with a width corresponding to an underlying gaussian function of 0.15 eV Full Width at Half Maximum (FWHM). Some curves have been shifted vertically to enhance the comparison.}
\label{ODOS}
\end{figure}

\begin{table}[t]
\caption{5f occupancies and branching ratios of the selected actinides. For $\delta$-Pu, the results corresponding to the reduced muffin-tin (MT) radius (equal to the muffin-tin radius for $\alpha$-U) are given to illustrate the dependence of occupation numbers on the details of MT-geometry. Branching ratios were evaluated following the formula in Eq. (\ref{eqn10}): $BR=1-\frac{14}{15}\frac{N_{5/2}}{N}$ where $N_{5/2}$ is the number of $5f_{5/2}$ holes and $N$ is the total number of $5f$ holes. Experimental XAS and EELS results are cited from Ref. [\onlinecite{prb_72_085109}].} \label{br_5f}
\small
\begin{center}
\begin{tabular}{@{}c c c c c} &$n_{5/2}$&$n_{7/2}$&$n_{5/2}+n_{7/2}$&BR\\
\hline\hline
$\alpha$-U& & &&\\
LDA&1.448 &1.078 &2.525&0.630\\
scGW&1.269 &0.946 &2.215&0.625\\
sc(GW+G3W2)&1.301 &0.957 &2.258&0.626\\
XAS& & &&0.676\\
EELS& & &&0.685\\
$\delta$-Pu ($R_{MT}=3.0965$)& & &&\\
LDA&4.101 &0.929 &5.03&0.802\\
scGW&4.769 &0.250 &5.019&0.872\\
sc(GW+G3W2)&4.569 &0.340 &4.909&0.853\\
XAS& & &&0.813\\
EELS& & &&0.826\\
$\delta$-Pu ($R_{MT}=2.602$)& & &&\\
LDA&3.893 &0.825 &4.718&0.788\\
scGW&4.391 &0.217 &4.608&0.840\\
\end{tabular}
\end{center}
\end{table}

Our the most sophisticated theoretical approach which we use in this work is sc(GW+G3W2). Before comparing the experimental results and the results obtained with this approach, let us discuss the differences in the electronic structure obtained with sc(GW+G3W2) and two other theoretical approximations we use: LDA and scGW. Figure \ref{DOS_theor} shows total DOS of $\alpha$-U and $\delta$-Pu as obtained in calculations. The most striking fact which one can see in Figure \ref{DOS_theor} is that all three methods result in very similar spectra for $\alpha$-U but they differ a lot when applied to $\delta$-Pu. A direct consequence of this fact is that differences in treatment of exchange-correlation effects between LDA, scGW, and sc(GW+G3W2) are not essential in the case of $\alpha$-U. At the same time, the way one treats these effects makes a big difference in the case of $\delta$-Pu. Particularly, even correlation effects beyond scGW approximation are rather strong which can be evidenced by looking at sc(GW+G3W2) DOS. We will use this observation when we compare theoretical results with experimental data.

One more interesting thing which can be seen in Fig. \ref{DOS_theor} for $\alpha$-U is the sub-structure of the second peak positioned at 1.5 eV. In LDA, it has well defined two-lobe sub-structure. In scGW and especially in sc(GW+G3W2), the sub-structure almost disappears and becomes represented by one peak and one shoulder. Again, we will return to this point when compare our results with BIS spectra.

Figure \ref{PDOS_theor} presents partial density of states (PDOS) of two materials as obtained in calculations performed at three levels of approximation. Similar to the total DOS, we do not see noticeable differences between LDA, scGW, and sc(GW+G3W2) in the structure and relative positions of $5f_{5/2}$ and $5f_{7/2}$ peaks in $\alpha$-U. However, in $\delta$-Pu the differences are remarkable. Whereas one can see an appreciable mixing of $5f_{5/2}$ and $5f_{7/2}$ states in LDA, bringing non-local self-energy effects into consideration (scGW and sc(GW+G3W2)) makes the corresponding mixing almost negligible. In other words, non-local self-energy effects enhance spin-orbit splitting in $\delta$-Pu. As it was already shown in total DOS (Fig. \ref{DOS_theor}), the effect of the correlations beyond GW approximation consists simply in shifting the $5f_{7/2}$ peak to the lower (as compared to scGW) energy.

Now we turn to the principal part of our comparison. In Fig. \ref{UDOS} we compare UDOS of $\alpha$-U and $\delta$-Pu as obtained in sc(GW+G3W2) approximation with experimental IPES and BIS ($\alpha$-U only) spectra. Note a remarkable quantitative agreement of theoretical UDOS and experimental BIS spectrum for $\alpha$-U. The agreement is almost perfect for both the relative intensity of two peaks ($5f_{5/2}$ and $5f_{7/2}$) and their energy positions. Note also the absence of two-lobe sub-structure in the $5f_{7/2}$ peak of the BIS spectrum which supports its diminishing in sc(GW+G3W2) as compared to the LDA case mentioned before.

IPES spectrum for $\alpha$-U shows qualitative but only semi-quantitative agreement with BIS and sc(GW+G3W2) results. Namely, the position of the first peak (0.5--1 eV range) is in a good agreement with other spectra. But the energy position of the second peak seems to be underestimated. Also, the intensity of the second peak is rather low in IPES whereas in BIS and in the calculations the intensity of the second peak is higher than of the first one. Possible reasons for disagreements of IPES and BIS spectra for light actinides were discussed by P. Roussel et al. in [\onlinecite{ss_714_121914}]. We can add here that the 6d:5f cross section ratio at 10 eV (IPES) is about 500 times larger than at 1486 eV (BIS). This could be the reason that 5f peaks are less pronounced in IPES. Thus, a logical conclusion about UDOS of $\alpha$-U which one can draw from our work is that BIS spectrum is close to the true UDOS of $\alpha$-U. This conclusion is supported by all three our theoretical calculations which also agree between each other. Deviation of IPES result from BIS is most likely related to the high 6d:5f cross section in IPES experiments.

UDOS of $\delta$-Pu has one peak which is associated with $5f_{7/2}$ states. As one can conclude from Fig. \ref{UDOS}, theoretical result (position of the peak) is shifted slightly towards higher energies as compared to the IPES peak position. However, the shift is within the IPES energy resolution (about 1 eV). There are some possible explanations for the reason of the shift as well. Firstly, as it is evident from Fig. \ref{DOS_theor}, the addition of the first order vertex corrections to scGW approximation which is embodied in sc(GW+G3W2) method moves the position of $5f_{7/2}$ peak towards lower energies by approximately 1.2 eV. Therefore, it is very likely that if one adds more diagrams (i.e. those beyond sc(GW+G3W2)) the $5f_{7/2}$ peak will be moved further to the left (but the corresponding shift would be smaller, maybe about 0.5 eV). Secondly, as in many other situations such as, for instance, band gaps in semiconductors/insulators\cite{prb_95_195120}, LDA and sc(GW+G3W2) provide lower and upper limits of the exact result. Therefore, from this point of view, if one takes an average of LDA and sc(GW+G3W2) peak positions the result also will be at about 1.9--2 eV in good agreement with IPES. Thirdly, if higher order diagrams (beyond sc(GW+G3W2)) have little effect on the position of $5f_{7/2}$ peak, there is a possibility that its position is underestimated in IPES experiments, similar to the differences between IPES and BIS for $\alpha$-U.

\begin{figure}[t]    
    \includegraphics[width=7 cm]{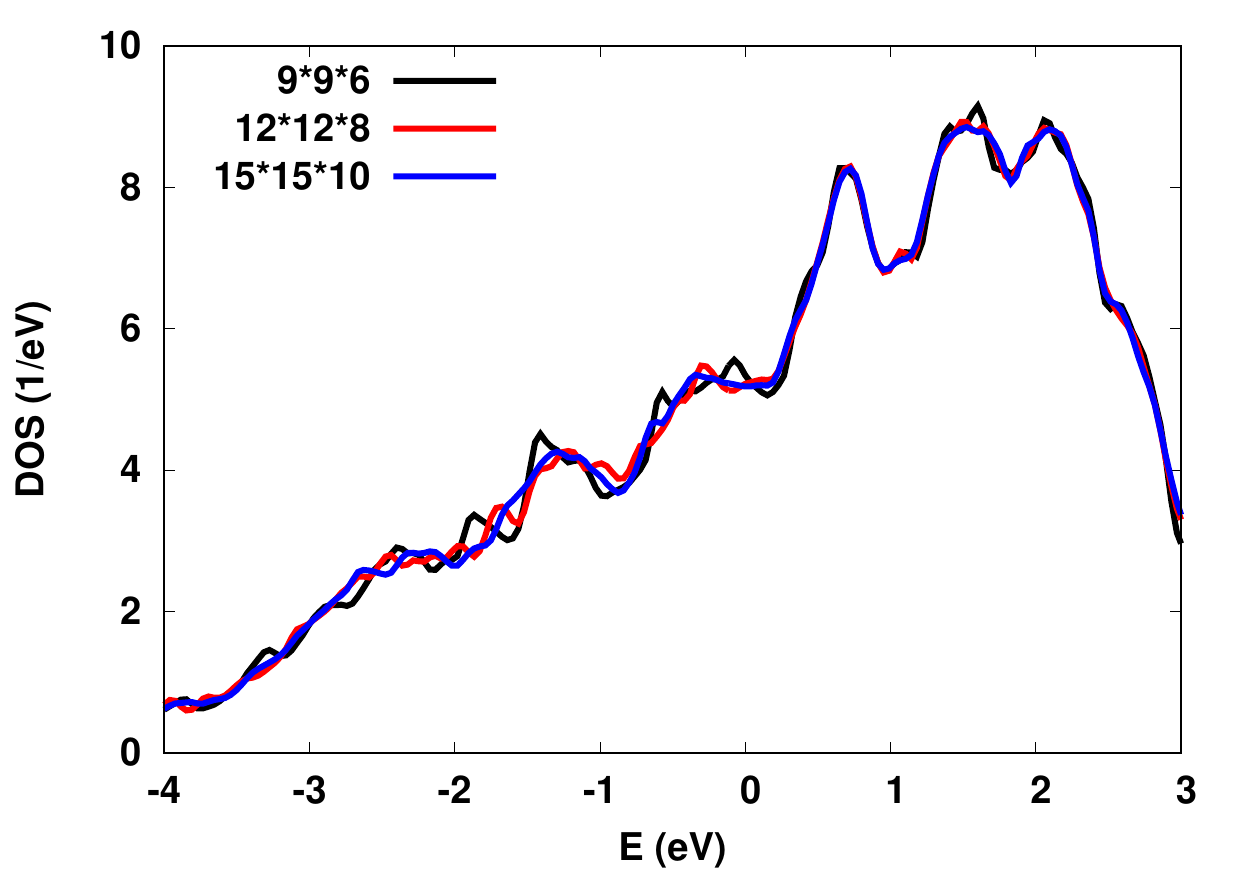} 
\caption{DOS of $\alpha$-U obtained in LDA calculations with different \textbf{k}-grids. Note gradual disappearance of the fine structure between -0.5 eV and the Fermi level.}
\label{DOS_k}
\end{figure}

Let us consider now the occupied density of states presented in Fig. \ref{ODOS}. Theoretical results for $\alpha$-U are compared with PES data obtained by Naegele, [\onlinecite{jnm_166_59}]. Essentially, there are three points of importance in the comparison. Firstly, principal peak right under the $E_{F}$ is represented well in the calculations. Secondly, at least some peaks visible in the calculations but not in PES, in fact, are the artifacts of the insufficient number of points in the Brillouin zone used for the integration. We checked this at the LDA level of approximation because scGW and sc(GW+G3W2) calculations are too time consuming for the denser \textbf{k}-grids than we used ($9\times 9\times 6$). The validity of the LDA checking (instead of direct checks of sc(GW+G3W2)) can be justified by noticing that ODOS of $\alpha$-U has sub-structures in the range between -0.5 eV and the Fermi energy in all theoretical variants. As can be seen from Fig. \ref{DOS_k}, when one increases the density of \textbf{k}-points the features right under the $E_{F}$ are washed out leaving only one principal peak right below the Fermi level (similar to the experimental PES spectrum). Thirdly, as one also can see from Fig. \ref{DOS_k}, the peak at about -1.5 eV does not disappear when the number of \textbf{k}-points increases. This peak seems to be a robust result in calculations. In this respect, PES shows a couple of rather small shoulders at about -1.1 eV and -2 eV. Whether one of them corresponds to the theoretical peak at -1.5 eV and if it is what is the reason for the difference in its position is not yet clear at this point.

Theoretical results for $\delta$-Pu are compared  with ResPES data obtained by J. Tobin et al., [\onlinecite{prb_68_155109}]. Similar to $\alpha$-U, sharp ResPES peak right below the $E_{F}$ is reproduced well in the calculations. Second principal peak of the ResPES spectrum - broad peak with the position of its maximum at about -0.9 eV is represented by a shoulder in the calculations (same energy position). In this respect, another experimental data, PES spectrum obtained by Havela et al.\cite{prb_65_235118}, shows this structure in closer correspondence to our calculations than to the ResPES spectrum. Namely, PES spectrum of $\delta$-Pu in Ref. [\onlinecite{prb_65_235118}] has a sharp peak right below the $E_{F}$ and a smaller feature (more like a shoulder) at -0.9 eV. Thus, in general, ODOS of $\delta$-Pu obtained from sc(GW+G3W2) calculations is in a good agreement with the experimental spectroscopy.

Table \ref{br_5f} presents 5f occupancies as well as branching ratios. Before comparing the calculated results with experimental data, however, let us make an important remark about muffin-tin (MT) geometry effect. Occupation numbers which we present in the table correspond to only the volume inside of the MT spheres which surround atoms. MT spheres do not overlap and, correspondingly, their volume depends on particular atomic arrangement, i. e. on the crystal structure. For our analysis it is important that the volume of MT spheres in $\delta$-Pu is almost twice larger than the volume of $\alpha$-U MT spheres (see the MT radii in Table \ref{list_s}). The part of $5f$ orbital which geometrically is outside of the MT spheres is represented by plane waves and, correspondingly, does not contribute to the occupation number. Therefore, the numbers presented in Table \ref{br_5f} are a bit smaller than the real occupation numbers which would appear in the calculations if the effect of MT geometry was not present. For $\delta$-Pu the volume of MT spheres is sufficiently large and the mismatch is expected to be small (probably less than 0.1) but for $\alpha$-U it is expected to be larger. In order to estimate the mismatch in the occupation numbers of $\alpha$-U we did a little numerical experiment: we evaluated occupations for $\delta$-Pu using two different MT radii: i) the biggest possible; ii) equal to the MT radius of $\alpha$-U. As one can see from Table \ref{br_5f}, the mismatch in the $5f$ occupancies is about 0.3 which, therefore, we consider as a possible uncertainty in the evaluated occupation numbers for $\alpha$-U. As a result, branching ratios for $\alpha$-U does not fit well to the experimental results. For $\delta$-Pu, however, comparison of the calculated branching ratios with the experimental values supports our speculation (see above) that LDA and sc(GW+G3W2) approximations provide lower and upper limits of the correct value. Also, the total $5f$ occupation is close to its experimental value (approximately 5) in all three theoretical methods (remember slight increase related to the MT geometry effect).

We also would like to briefly mention one aspect related to the electronic structure of actinides which is considered to be of importance currently. This is about the quantification of delocalization in the $5f$ states of the actinides. As it is suggested in Ref. [\onlinecite{applsci_11_3882}], the combination of XAS and XES can be very useful in this respect. Specifically, XAS is sensitive to the partial $5f_{5/2}$ and $5f_{7/2}$ UDOS whereas XES is sensitive to the corresponding ODOS. Therefore, their combination can help in the quantification of the mixing of $5f_{5/2}$ and $5f_{7/2}$ states, i.e. of the delocalization. In the next section we provide all the details of this new approach for the quantification of delocalization and we present first results of studies using $\alpha$-U, UO$_{2}$, and UF$_{4}$ as examples. It is important to mention it here because theoretical calculations can be of a help in this question. Particularly, we would like to point out that theoretical results for partial $5f_{5/2}$ and $5f_{7/2}$ occupations also are sensitive to the crystal structure/chemical environment in a specific compound. In order to illustrate this we have performed the calculations of occupation numbers in UO$_{2}$ and obtained the following values: $n_{5/2}+n_{7/2}=1.710+0.560$ (LDA) and $n_{5/2}+n_{7/2}=1.845+0.324$ (scGW). The calculations were a bit simplified (we used ferro-magnetic ordering instead of experimentally observed antiferromagnetic ordering) but they clearly show considerably smaller mixing of the occupied $5f_{5/2}$ and $5f_{7/2}$ states in UO$_{2}$ as compared to $\alpha$-U (see Table \ref{br_5f}). This is in accordance to experimental observations made in Ref. [\onlinecite{applsci_11_3882}] and in the following section. Therefore, as an interesting and, we hope, important prospect of ab-initio modelling would be to assist in the interpretation of XES studies of the delocalization effects.

\section*{X-Ray Emission Spectroscopy}
\label{xes}

Delocalization in the actinide 5f states is a very important phenomenon, but not very well understood.  The effect of 5f delocalization can be seen in one of the most fundamental of elemental parameters, atomic size.  In the early actinides, the Wigner-Seitz radii change with filling in a manner consistent with the addition of delocalized electrons.\cite{las_26_128,las_26_16,las_26_90} The observation of this effect was so striking that it lead temporarily to the incorrect hypothesis that the Actinides were a 6d, not 5f, filling series.\cite{jinc_35_3487} This misconception was subsequently corrected, as it was shown that 5f filling could account for the observed behavior.\cite{prl_41_42} Nevertheless, the measurement of 5f dispersions with angle-resolve photoelectron spectroscopy has been something of a disappointment.  While metallic U valence states can show energy variation with crystal momentum,\cite{prb_73_165109,prb_75_045120} the dispersion of the 5f derived states is very weak,\cite{applsci_11_3882} on the order of 0.1 eV, with little or no exhibition of connection to the high symmetry points or lines in the Brillouin Zone, both of which are easily seen in strongly dispersing systems.\cite{prb_28_6169,prb_32_3465,prb_33_2270,prb_35_9056,prb_45_5563} Moreover, it has been known for decades,\cite{prb_35_2667,prb_92_035111} that the X-ray Absorption Spectroscopy (XAS) Branching Ratio (BR) is the same for localized n= 2 systems (UF$_{4}$ and UO$_{2}$) and the delocalized, n = 3 system, metallic U. Thus, the situation is problematic: there should be strong 5f dispersion in the early actinides, but spectroscopically no manifestation can be found.

\begin{figure}[t]    
\includegraphics[width=7 cm]{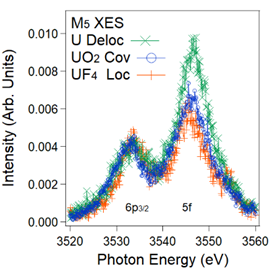} 
\caption{Here are shown the M$_{5}$ XES of UF4 (red +), UO2 (blue o) and U metal (green x).  The horizontal axis is emitted photon energy. Backgrounds have been subtracted and the spectra have been equated at the M$_{5}$ 6p3/2 peak height (hv ~ 3533 eV).  The M$_{5}$ 5f peaks are at hv ~ 3547 eV.  This is non-resonant XES, with the excitation energies slightly above threshold and out of the resonant regime. The 5f:6p Peak Ratios are 3.21 (U), 2.29 (UO$_{2}$) and 1.92 (UF$_{4}$), where all fits are with Lorentzians of FWHM(6p) = 7 eV and FWHM(5f) = 9 eV. FWHM is full width at half max.  Peak area   determinations were typically good to 1\%, generating a peak ratio error of about 2\%.}
\label{fig8}
\end{figure}

Here, that situation is rectified.  Shown in Figure \ref{fig8} are the U M$_{5}$ X-ray Emission Spectra (XES) for UF$_{4}$, UO$_{2}$ and U metal.  Clearly, the 5f delocalized U metal exhibits a substantially different spectrum than those for the localized UF$_{4}$ and UO$_{2}$ samples.  This is the behavior that has been expected but not previously observed.  There may even be a slight difference between the UF$_{4}$ and UO$_{2}$ cases, consistent with the interpretation that UF$_{4}$ is a simple, highly localized 5f system and UO$_{2}$ is a 5f localized but covalent case.\cite{prb_92_045130}

Below, it will be demonstrated that the change in the M$_{5}$ 5f peak intensity for metallic U is consistent with (1) electric dipole selection rules and transition moments and (2) delocalization effects relative to the Intermediate Coupling Model.\cite{prb_53_14458,prl_93_097401,prb_72_085109} To do that, the discussion will digress to a consideration of XAS BR measurements and then proceed with a consideration of XES, including the development of a peak ratio (PR) picture within the constraints of experimental results for the M$_{4}$ and M$_{5}$ XES of UF$_{4}$. Finally, including a correction for surface oxidation effects, the measured U metal XES will be compared to the predictions of the PR model.

\begin{table}[t]
\caption{U N$_{4,5}$ (4d) XAS Branching Ratios and 5f Populations.  The BR values are experimentally determined from XAS spectra.} \label{tbl4}
\small
\begin{center}
\begin{tabular}{@{}c c c c c c c c c c} &n&BR&$n_{5/2}$&$n_{7/2}$&$N$&$N_{5/2}$&$N_{7/2}$&$N_{5/2}/N$&$N_{7/2}/N$\\
\hline\hline
Interm. coupl.,&2&0.68 &1.96 &0.04&12  &4.04&7.96&0.337&0.663\\
UO$_{2}$ and UF$_{4}$&& & &&  &&&$\sim$0.34&$\sim$0.66\\
\hline
U metal&3&0.68 &2.23 &0.77&11  &3.77&7.23&0.343&0.657\\
  && & &&  &&&$\sim$0.34&$\sim$0.66\\
\end{tabular}
\end{center}
\end{table}

A key issue discussed above, is that the XAS Branching Ratios for the localized U samples (UO$_{2}$ and UF$_{4}$) is the same as that for the delocalized U (U metal).  It turns out that this should not be a surprising result.  It will be shown below that the BR values depend solely upon the percentage of the unoccupied 5f states, that is either $5f_{5/2}$ or $5f_{7/2}$.  As can be seen in Table \ref{tbl4}, the percentage unoccupations are the same for all three samples.  Here, N, N$_{5/2}$ and N$_{7/2}$ are the total number of 5f holes, the number of $5f_{5/2}$ holes and the number of $5f_{7/2}$ holes, respectively.  Obviously, $N_{5/2} + N_{7/2} = N$.  For UO$_{2}$ and UF$_{4}$, $N = 14 -2 =12$ and for U metal $N = 14 -3 = 11$.\cite{prb_92_035111,applsci_10_2918,naegele94}

\begin{table}[b]
\caption{The relative 5f electric dipole cross sections are shown here.  See Refs. [\onlinecite{jpc_4_015013}] and [\onlinecite{applsci_10_2918}] for details.  Note that the relative 5f cross section total = 28/3.   The 7/2 - 3/2 transition is forbidden.} \label{tbl5}
\small
\begin{center}
\begin{tabular}{@{}c c c c} &&$5f_{5/2}$&$5f_{7/2}$\\
&&Empty (Full)&Empty (Full)\\
\hline\hline
$N_{5}$&$d_{5/2}$&4/15&16/3\\
($M_{5}$)&Full (Empty)&&\\
\hline
$N_{4}$&$d_{3/2}$&56/15&0\\
($M_{4}$)&Full (Empty)&&\\
\end{tabular}
\end{center}
\end{table}

To understand this, first the electric dipole transitions for the $f\rightarrow d$ case must be obtained.  This is a fairly simple procedure and is discussed in detail elsewhere.\cite{jpc_4_015013,applsci_10_2918} The results are shown in Table \ref{tbl5}.  For the XAS BR, the transitions are from the 4d states into the empty 5f states, as summarized in Equations (\ref{eqn6}) and (\ref{eqn7}).  Historically, the great success of the BR within the Intermediate Coupling Model\cite{prb_92_035111,prl_93_097401,prb_72_085109} indicates that the electric dipole selection rules and cross sections must be accurate.  One aspect of this is that the $4d_{3/2}$ to $5f_{7/2}$ transition is forbidden.
\begin{equation}\label{eqn6}
4d_{3/2}+h\nu\rightarrow 5f_{5/2},
\end{equation}
\begin{equation}\label{eqn7}
4d_{5/2}+h\nu\rightarrow 5f_{5/2}+5f_{7/2}.
\end{equation}

Of course, the 5f states are not completely empty.  Using the cross sections in Table \ref{tbl5} and the percentage unoccupations, $N_{5/2}/N$ and $N_{7/2}/N$, it is possible to predict the relative intensities and branching ratio, as illustrated in Equations (\ref{eqn8})-(\ref{eqn10}).\cite{jesrp_232_100,ss_698_121607}
\begin{equation}\label{eqn8}
I_{5/2}=\frac{N_{5/2}}{6}\Big(\frac{4}{15}\Big)+\frac{N_{7/2}}{8}\Big(\frac{80}{15}\Big),
\end{equation}
\begin{equation}\label{eqn9}
I_{7/2}=\frac{N_{5/2}}{6}\Big(\frac{56}{15}\Big),
\end{equation}
\begin{align}\label{eqn10}
BR=\frac{I_{5/2}}{I_{5/2}+I_{7/2}}&=\Big(\frac{1}{15}\Big)\Big(\frac{N_{5/2}}{N}\Big)+\Big(\frac{N_{7/2}}{N}\Big)\nonumber\\&=1-\Big(\frac{14}{15}\Big)\Big(\frac{N_{5/2}}{N}\Big),
\end{align}

If Equation (\ref{eqn8}) is applied to the percentage unoccupations for various samples and models\cite{prb_92_035111,prl_93_097401,prb_72_085109} it will be seen that it is completely accurate.  Next, a parallel analysis will be applied to XES.

\begin{figure}[t]    
\includegraphics[width=7 cm]{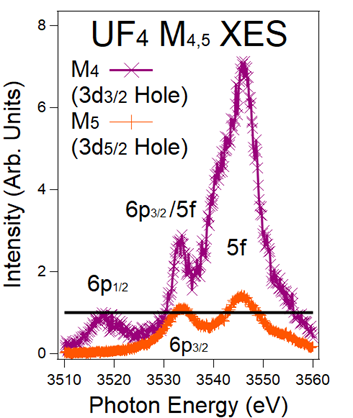} 
\caption{The $M_{4}$ and $M_{5}$ XES spectra of UF$_{4}$ are shown here. The normalization of the spectra is via the 6p peaks: $M_{4}$ $6p_{1/2}$ ($p_{1/2}\rightarrow d_{3/2}$) and $M_{5}$ $6p_{3/2}$ ($p_{3/2}\rightarrow d_{5/2}$), both $\Delta j = 1$.  The black horizontal line is at unity (1).  The intensity ratio of the $6p_{1/2} : 6p_{3/2}$ is 0.8, following the electric dipole cross sections.  In order to align the peaks on the $M_{5}$ energy scale, the $M_{4}$ spectrum has been shifted -181 eV.  The details of the normalization are available in Refs. [\onlinecite{phsynres_102_241}] and [\onlinecite{matscieng_109_012007}].  The $6p_{1/2}$ to $3d_{5/2}$ transition is electric dipole forbidden and thus absent in the $M_{5}$ spectrum.}
\label{fig9}
\end{figure}

Now, consider the experimental results for the M$_{4,5}$ ($3d$) XES of UF$_{4}$, shown in Figure \ref{fig9}.  As discussed in detail elsewhere,\cite{phsynres_102_241,matscieng_109_012007} these spectra clearly indicate that both (1) the electric dipole selection rules and cross sections and (2) the Intermediate Coupling Model apply for U M$_{4,5}$ XES.  However, unlike the $4d$ XAS BR measurements, here the normalization is through the $6p$ XES and, as might be expected, there will need to be a higher order correction in the cross-section analysis.\cite{jpc_4_015013,applsci_10_2918}

\begin{table}[t]
\caption{The $5f$ Electric dipole cross sections are shown here.  The total $5f$ cross section is 28/3.  The $5f$ cross section per $3d$ hole is 14/15.  Left (right) are the results corresponding to completely empty $3d$ states (per $3d$ hole).  See Refs. [\onlinecite{jpc_4_015013}] and [\onlinecite{applsci_10_2918}] for details.} \label{tbl6}
\small
\begin{center}
\begin{tabular}{@{}c c c c c c c} &&$5f_{5/2}$&$5f_{7/2}$&$5f_{5/2}$&$5f_{7/2}$\\
&&Empty&Empty&&Full&Full\\
&&(Full)&(Full)&&Full&Full\\
\hline\hline
$N_{5}$&$d_{5/2}$&4/15&16/3&$d_{5/2}$&4/90&16/18\\
($M_{5}$)&Full&&&1&&\\
&(Empty)&&&Hole&&\\
\hline
$N_{4}$&$d_{3/2}$&56/15&0&$d_{3/2}$&56/60&0\\
($M_{4}$)&Full&&&1 &&\\
&(Empty)&&&Hole &&\\
\end{tabular}
\end{center}
\end{table}

To begin, once again the electric dipole cross sections are required.  These are shown in Table \ref{tbl6}.  For XES, it is necessary to normalize the cross sections per $3d$ hole.  Using the cross sections in Table \ref{tbl6} and the percentage occupations, $n_{5/2}/n$ and $n_{7/2}/n$, it is possible to predict the relative intensities and XES peak ratios, as illustrated in Equations (\ref{eqn11})-(\ref{eqn13}). Here, $n$, $n_{5/2}$ and $n_{7/2}$ are the total number of $5f$ electrons, the number of $5f_{5/2}$ electrons and the number of $5f_{7/2}$ electrons, respectively with $n_{5/2} + n_{7/2} = n$.

\begin{equation}\label{eqn11}
I_{5f}^{N_{5}}=\frac{n_{5/2}}{6}\Big(\frac{4}{90}\Big)+\frac{n_{7/2}}{8}\Big(\frac{16}{18}\Big),
\end{equation}
\begin{equation}\label{eqn12}
I_{5f}^{N_{4}}=\frac{n_{5/2}}{6}\Big(\frac{56}{60}\Big),
\end{equation}

\begin{equation}\label{eqn13}
PR_{5f}^{N}=\frac{I_{5f}^{N_{4}}}{I_{5f}^{N_{5}}}=\frac{\frac{n_{5/2}}{6}\Big(\frac{56}{60}\Big)}{\frac{n_{5/2}}{6}\Big(\frac{4}{90}\Big)+\frac{n_{7/2}}{8}\Big(\frac{16}{18}\Big)}=\frac{21n_{5/2}}{n_{5/2}+15n_{7/2}}.
\end{equation}

For the $N (4d)$ transitions, one might expect that the electric dipole selection rules and cross sections would hold almost perfectly, as they do for the 4d XAS BR.  However, for the $M (3d)$ transitions, the energies are higher and the possibility of higher order terms increases.\cite{jpc_4_015013,applsci_10_2918} Thus in Eq. (\ref{eqn14}), a correction term is included in the denominator. Based upon Figure \ref{fig9} and the more extensive analysis of Refs. [\onlinecite{jpc_4_015013,applsci_10_2918}], the PR $\sim$5. If a = 0 or Eq. (\ref{eqn13}) were to be used, the ratio should be in the range of 16 to 21. (PR  = 16 for $n_{5/2}$ = 1.96 and $n_{7/2}$ = 0.04 or PR = 21 for $n_{5/2}$ = 2 and $n_{7/2}$ = 0.) Of course, selection rule and cross section breakdowns are at their worst when cross sections are small, as is the case in the denominators of Eq. (\ref{eqn13}) and (\ref{eqn14}). Thus the correction term on Eq. (\ref{eqn14}) is not unexpected or unreasonable.  This analysis is discussed in detail in Refs. [\onlinecite{applsci_11_3882}] and [\onlinecite{applsci_10_2918}].

\begin{equation}\label{eqn14}
PR_{5f}^{M}=\frac{I_{5f}^{M_{4}}}{I_{5f}^{M_{5}}}=\frac{21n_{5/2}}{n_{5/2}+15n_{7/2}+a}.
\end{equation}

\begin{table}[t]
\caption{Shown here are the $6p$ electric dipole cross sections.  The total $6p$ cross section is 4.  The $6p$ cross section per $3d$ hole is 2/5.  Left (right) are the results for completely empty $3d$ states (per $3d$ hole).} \label{tbl7}
\small
\begin{center}
\begin{tabular}{@{}c c c c c c c} &&$6p_{1/2}$&$6p_{3/2}$&&$6p_{1/2}$&$6p_{3/2}$\\
&&Full&Full&&Full&Full\\
\hline\hline
$M_{5}$&$3d_{5/2}$&0&12/5&$3d_{5/2}$&0&2/5\\
&Empty&&&1 Hole&&\\
\hline
$M_{4}$&$d_{3/2}$&4/3&4/15&$d_{3/2}$&1/3&1/15\\
&Empty&&&1 Hole &&\\
\end{tabular}
\end{center}
\end{table}

Experimentally, the normalization is through the $6p$ peaks.  Thus, it is more effective to couch the normalization analysis in terms of the $6p$ peaks.  To do this, the $6p$ cross sections per $3d$ hole are required.  These are shown in Table \ref{tbl7}.  Note that the electric dipole selection rules hold better for the $6p\rightarrow 3d$ transitions than the $5f\rightarrow 3d$.  This can be seen in the absence of the $6p_{1/2}$ peak from the M$_{5}$ XES spectrum in Figure \ref{fig8}.  Further detail can be found in Refs. [\onlinecite{applsci_11_3882}] and [\onlinecite{prl_93_097401}].

Using the $6p$ cross sections and the experimental results in Figure \ref{fig9}, it is possible to obtain the relationships in Eq. (\ref{eqn15}) and (\ref{eqn16}). As can be seen in Table \ref{tbl5}, the $n_{5/2}$ values for the localized cases (UO$_{2}$ and UF$_{4}$) and delocalized case (U metal) are only ~10\% different.  Thus, experimentally, the effect in the M$_{4}$ spectrum is expected to be small.  Fortunately, for the M$_{5}$ spectrum and Eq. (\ref{eqn16}), the situation is very different.  With its strong dependence upon the $n_{7/2}$ population, here is where the change from localized to delocalized should be most obvious.  This is what is observed in Figure \ref{fig8}.

\begin{equation}\label{eqn15}
PR_{fp}^{M_{4}}=\frac{I_{5f}^{M_{4}}}{I_{6p_{1/2}}^{M_{4}}}=5.31n_{5/2}.
\end{equation}
\begin{equation}\label{eqn16}
PR_{fp}^{M_{5}}=\frac{I_{5f}^{M_{5}}}{I_{6p_{3/2}}^{M_{5}}}=0.2098n_{5/2}+3.150n_{7/2}+1.361.
\end{equation}

However, the result is not quantitative.  Qualitatively, there is agreement between the M$_{5}$ XES and the prediction of Eq. (\ref{eqn16}).  To get quantitative agreement, surface oxidation must be addressed.

\begin{figure}[t]    
\includegraphics[width=7 cm]{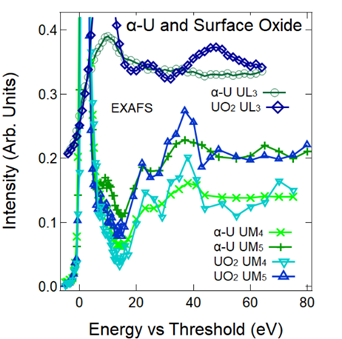} 
\caption{The X-ray Absorption Spectroscopy (XAS), including EXAFS, of bulk U metal and bulk UO$_{2}$ samples, is illustrated here.  The L edge work shows no surface sensitivity but the M edge does.  See text and Ref. [\onlinecite{ss_698_121607}] for more detail.}
\label{fig10}
\end{figure}

Dealing with radioactive samples can be difficult, oftentimes requiring triple sample containment.\cite{prb_92_035111} Hence, the thrust away from lower energy experiments, which require ultra-high vacuum and the concomitant absence of the triple containment.\cite{prb_68_155109} While dealing with surface issues is a common topic in soft X-ray experiments\cite{prb_92_045130,jvsta_35_03E108}, usually in the harder X-Ray regime such considerations can be neglected.  Not surprisingly, in the Tender X-ray regime of the U M$_{4,5}$ XES experiments, there is an enhanced surface sensitivity relative to the L$_{3}$ experiments at 17 keV.  These effects were reported and discussed in detail in Ref. [\onlinecite{ss_698_121607}].  Examples of the results are shown in Figure \ref{fig10}.  Here the first EXAFS peaks, associated with the U-O bond, is used as a probe of oxidation.  (EXAFS is Extended X-ray Absorption Fine Structure, with features at 20 eV or more above the absorption edge or white line.) Clearly, the L$_{3}$ EXAFS shows no surface sensitivity to the oxide.  However, both the M$_{4}$ and M$_{5}$ EXAFS show the beginnings of the UO$_{2}$ peak, and a sensitivity to surface oxidation.  It appears that the EXAFS peak is about 1/2 of the value corresponding to pure UO$_{2}$, thus it is estimated that 1/2 of the signal in the U M XES may be coming from UO$_{2}$ instead of U metal.

\begin{figure}[t]    
\includegraphics[width=7 cm]{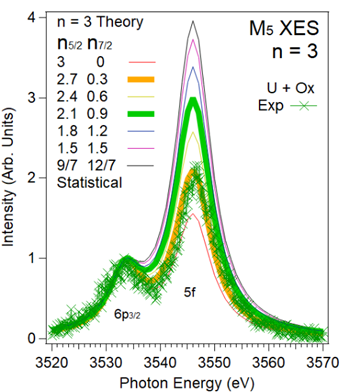} 
\caption{The experimental M$_{5}$ XES spectrum of U metal (green +) and the simulated spectra for $n = 3$ and various values of $n_{7/2}$ are shown here.  The normalization is via the $6p$ peak. The simulation used normalized Lorentzian line shapes, with a half-width at half-max of $\Gamma$.  For the $6p_{3/2}: \Gamma = 3.5$ eV.  For the $5f$ peak, $\Gamma$ = 4 eV. All peaks were normalized at $6p_{3/2}$ to 1.  For metallic U, $n_{5/2} = 2.2$ and $n_{7/2} = 0.8$, from Refs. [\onlinecite{prb_53_14458}] and [\onlinecite{prl_93_097401}].  Because of their importance, a larger linewidth has been used for $n_{7/2}$ = 0.9 and 0.3.}
\label{fig11}
\end{figure}

At this point, it is possible to proceed to the quantitative comparison of the experiment and theory.

Using Eq. (\ref{eqn16}), it is possible to generate simulated spectra, as described in detail in Ref. [\onlinecite{applsci_10_2918}]. The results of this operation can be seen in Figure \ref{fig11}.  Along with the simulated spectra are the corresponding experimental data for U metal.  Based upon this comparison, the projected $n_{7/2}$ population of the U metal sample would be $\sim$0.3.  The analyses founded upon the Intermediate Coupling Model and the $4d$ BR results would predict a value of $\sim$0.8. However, as has been seen in the section above, it is estimated that about 1/2 of the XES signal may be coming from the oxidized surface.  Thus, within this caveat, there is essentially quantitative agreement between the predictions of the simple theory presented here and the experimental measurement of the U M$_{5}$ XES.  At worst, it should be noted that while quantifying the exact values at this stage is difficult, it is clear that the data and model agree qualitatively in any case.

\section*{Conclusions}
\label{concl}

In our ab-initio theoretical calculations we have applied (for the first time) self-consistent vertex corrected GW approximation, sc(GW+G3W2), to study the electronic structure of actinides $\alpha$-U and $\delta$-Pu. It has been shown that combined inclusion of relativistic effects (through Dirac's equation) and of the correlation effects (through the vertex corrected GW approach) allows one to describe ODOS/UDOS of two actinide metals in very good agreement with experimental ResPES/PES and BIS/IPES studies. Our ab-initio results allowed us to understand better subtle differences in experimental BIS and IPES spectra of $\alpha$-U. We also have suggested a future prospect of ab-initio assistance in new experimental XAS/XES studies of delocalization in uranium and its compounds which represents considerable interest in actinide community.

Several key milestones were achieved in this work from experimental point of view.  First, the anomaly of the identical U N$_{4,5}$ XAS Branching Ratio (BR) values for localized, $n = 2$, U (UO$_{2}$ and UF$_{4}$) and delocalized, $n = 3$, metallic U has been explained.\cite{jesrp_194_14} This was achieved by demonstrating that the XAS BR depends not upon percentage occupation but rather percentage unoccupation of the U $5f$ levels.  Second, Intermediate Coupling of the U $5f$ states was experimentally observed in XES.  Third, a parallel theoretical Peak Ratio (PR) framework for X-ray Emission Spectroscopy was developed and tested.  For the $M_{4,5}$ XES, the peak normalization is done through the spectator $6p$ peaks. Importantly, the PR depends not upon the unoccupation but rather the occupation of the U $5f$ levels.  Fourth, the effects of delocalization were measured in a metallic U sample and correlated with the PR picture, in connection with the framework of the intermediate Coupling Model.  Within a correction for surface oxidation, quantitative agreement was obtained between the PR predictions and the experimental measurements.

These results are important on a broader scale.  The XAS BR, coupled to the Intermediate Coupling Model, has had great success in the analysis of $5f$ localized actinide systems.  The most important aspect of this success was the transparent determination of the $5f$ occupation.  However, delocalized systems in particular and mixed systems in general remained problematic.  With the advent of the XES PR analysis, it will be possible to directly test for mixing between the $5f_{5/2}$ and $5f_{7/2}$ manifolds, away from the values predicted by the Intermediate Coupling Model.  In the case of U metal, it is understood that the mixing is driven by delocalization.  However, other types of mixing are expected, such as that from magnetic perturbations and electron correlation effects.  The application of XES to such systems should be highly illuminating.  For example, it should help to resolve some controversies surrounding URu$_{2}$Si$_{2}$ and related systems.\cite{prb_94_045121}

\section*{Acknowledgments}
\label{acknow}

Stanford Synchrotron Radiation Light-source is a national user facility operated by Stanford University on behalf of the DOE and the OBES. Part funding for the instrument used for this study came from the U.S. Department of Energy, Office of Energy Efficiency and Renewable Energy, Solar Energy Technology Office BRIDGE Program. Resources of the National Energy Research Scientific Computing Center, a DOE Office of Science User Facility supported by the Office of Science of the U.S. Department of Energy under Contract No. DE-AC02-05CH11231, were used in this work. LLNL is operated by Lawrence Livermore National Security, LLC, for the U.S. Department of Energy, National Nuclear Security Administration, under Contract DE-AC52-07NA27344. The work of AK was supported by the U.S. Department of energy, Office of Science, Basic Energy Sciences as a part of the Computational Materials Science Program.


\end{document}